\documentclass[%
 aip,
 jmp,%
 amsmath,amssymb,
 reprint,%
]{revtex4-1}

\usepackage[T1]{fontenc}
\usepackage{float}
\usepackage{textcomp}
\usepackage{amsmath}
\usepackage{cancel}
\PassOptionsToPackage{normalem}{ulem}

\usepackage{ulem}
\usepackage{color}
\usepackage{graphicx}
\usepackage{dcolumn}
\usepackage{bm}
\usepackage[mathlines]{lineno}
\usepackage{url}
\usepackage{hyperref}

\usepackage{epsfig}
\usepackage{epstopdf}
\usepackage{amsthm}
\usepackage{subfigure}

\definecolor{OliveGreen}{rgb}{0,0.6,0}
\definecolor{applegreen}{rgb}{0.55,0.71,0.0}
\definecolor{darkpastelgreen}{rgb}{0.01,0.75,0.24}
\definecolor{sandmaroon}{rgb}{0.7,0.3,0.2}   
\definecolor{orange}{rgb}{1.0,0.5,0.3}

\begin{document}

\preprint{AIP/123-QED}

\title{Binding  branched and linear DNA structures: from
isolated clusters to fully bonded gels}

\author{{J. Fernandez-Castanon}}
 \email{javier.fernandez.castanon@roma1.infn.it}
 \affiliation{Physics Department, Sapienza--Università di Roma, P.le A. Moro 5, 00185 Rome, Italy}
 
 \author{{F. Bomboi}}
\affiliation{Physics Department, Sapienza--Università di Roma, P.le A. Moro 5, 00185 Rome, Italy}

\author{F. Sciortino }
\affiliation{Physics Department,  Sapienza--Università di Roma, P.le A. Moro 5, 00185 Rome, Italy}
\affiliation{CNR-ISC, UOS Sapienza--Università di Roma, 00186 Rome, Italy}

\date{\today}

\begin{abstract}

The proper design of DNA sequences allows for the formation of well defined 
supramolecular units with controlled interactions via a consecution
of self-assembling processes.  Here, we  benefit from the controlled DNA self-assembly  
to experimentally realize   particles with well defined valence, namely
 tetravalent \emph{nanostars ($A$)} 
and bivalent \emph{chains ($B$)}. We specifically focus on the case in which 
$A$ particles can only bind to $B$ particles, via appropriately designed sticky-end sequences.  
Hence  \emph{AA}  and \emph{BB} bonds are not allowed.
Such a binary mixture system reproduces with DNA-based particles the physics of poly-functional condensation,
 with an exquisite control over the bonding process, tuned by the
 ratio, $r$, between $B$ and $A$ units  and by the temperature, $T$. We report dynamic light scattering experiments in a window of $T$s ranging from $10$\textdegree C to $55$\textdegree C and an interval of $r$ around the percolation transition to quantify the decay of the density correlation for the different cases.   At low $T$, when all possible bonds are formed,
the system behaves as a fully bonded network, as a percolating gel and as a cluster fluid depending on the selected $r$.

%
\end{abstract}

\maketitle

\section{\label{sec:Introduction}Introduction}  

Starting from the seminal work of Seeman~\cite{seeman1998dna}, several complex all DNA-made nanoparticles have been designed and
realized in the laboratories.  These novel colloidal particles have potential applications in several fields due to  their ability to operate at the nanoscale~\cite{bashir2001invited,pinheiro2011challenges}.  The possibility  to selectively control the binding of
complementary sequences and the different pairing ability of strands with different number of nucleotides are key elements in DNA design.  

Beside technological applications, DNA-made particles can be tailored to convert {\it in charta} or {\it in silico} intuitions into experimental realizations~\cite{biffi2013phase,bomboi2016re}. Indeed, it is not only 
possible to design the particles and their relative interactions with exquisite precision, but it is also feasible, exploiting self-assembly, to create them in bulk quantities and investigate their collective behavior.  One interesting field of application is the physics of patchy colloids,
particles interacting with a limited number of neighbours due to selective lock-and-key type binding.
These particles have stimulated a broad 
research for the peculiarities of their phase diagrams and unconventional physical properties~\cite{russo2011reentrant,de2011phase,de2012bicontinuous,
zhang2005self,smallenburg2013liquids,roldan2013gelling,duguet2016patchy}. 
The performance of these patchy-systems is strongly related to the number of patches present on particles' surfaces. In the single-bond-per-patch condition, the number of patches 
determines the maximum number of bonds that particles are able to form and  thus 
it is also referred to as  valence $f$.    For example,
theoretical studies suggest that $f$  controls the width of the
gas-liquid phase separation~\cite{bianchi2006phase,sciortino2017equilibrium}. 
 By reducing $f$,
 the phase separation boundary shifts to   small densities, opening a density region  in which,  at low enough temperatures ($T$), an $"empty"$ homogeneous arrested state, or $gel$, is formed~\cite{bianchi2006phase}.  Interestingly,  it has been recently suggested that 
the evolutionary process has  exploited the different gelation ability of proteins with  small valence to construct  self-assembled  biologically-relevant optical devices~\cite{cai2017eye}.
Beside proteins, concepts developed for patchy colloidal particles also apply~\cite{C7CP03149A} to telechelic and  poly-functional polymers, carrying complementary end-groups,  which have been investigated with the aim of controlling the strength of the resulting gel~\cite{sakai2008design,li2017sans}.

\smallskip{}

Previous experimental studies have shown  the 
suitability of DNA to produce bulk quantities of  particles in the lab to investigate the influence of   $f$ on the phase diagram of patchy colloids 
~\cite{biffi2013phase,bomboi2015equilibrium}.   Moreover,  experimental studies of binary mixtures of DNA particles of valence one and valence four  have been able to
reproduce with high fidelity theoretical predictions~\cite{roldan2013gelling} even for unconventional re-entrant gels, capitalizing on the competition between different bonding sequences~\cite{bomboi2015equilibrium}.
Here, we demonstrate how to design and implement an all-DNA system that constitutes the colloidal analog of poly-functional condensation but with the additional control of thermal reversibility and/or bond lifetime. This is a desirable step in the design of  biocompatible systems with innovative dynamic properties such as DNA-{\it vitrimers}~\cite{romano2015switching},  networks with self-healing properties~\cite{montarnal2011silica, denissen2016vitrimers}.
It also constitutes an advantageous stage for testing recent predictions on the long-range
nature of the effective interaction between colloidal particles immersed
in a solution which is itself close to percolation~\cite{gnancasimir2014}.  For this case,
the ability to control the lifetime of the clusters is crucial~\cite{gnan2012properties}.

\begin{figure}[ht] 
   \centering
   \includegraphics[width=3.5in, angle =0]{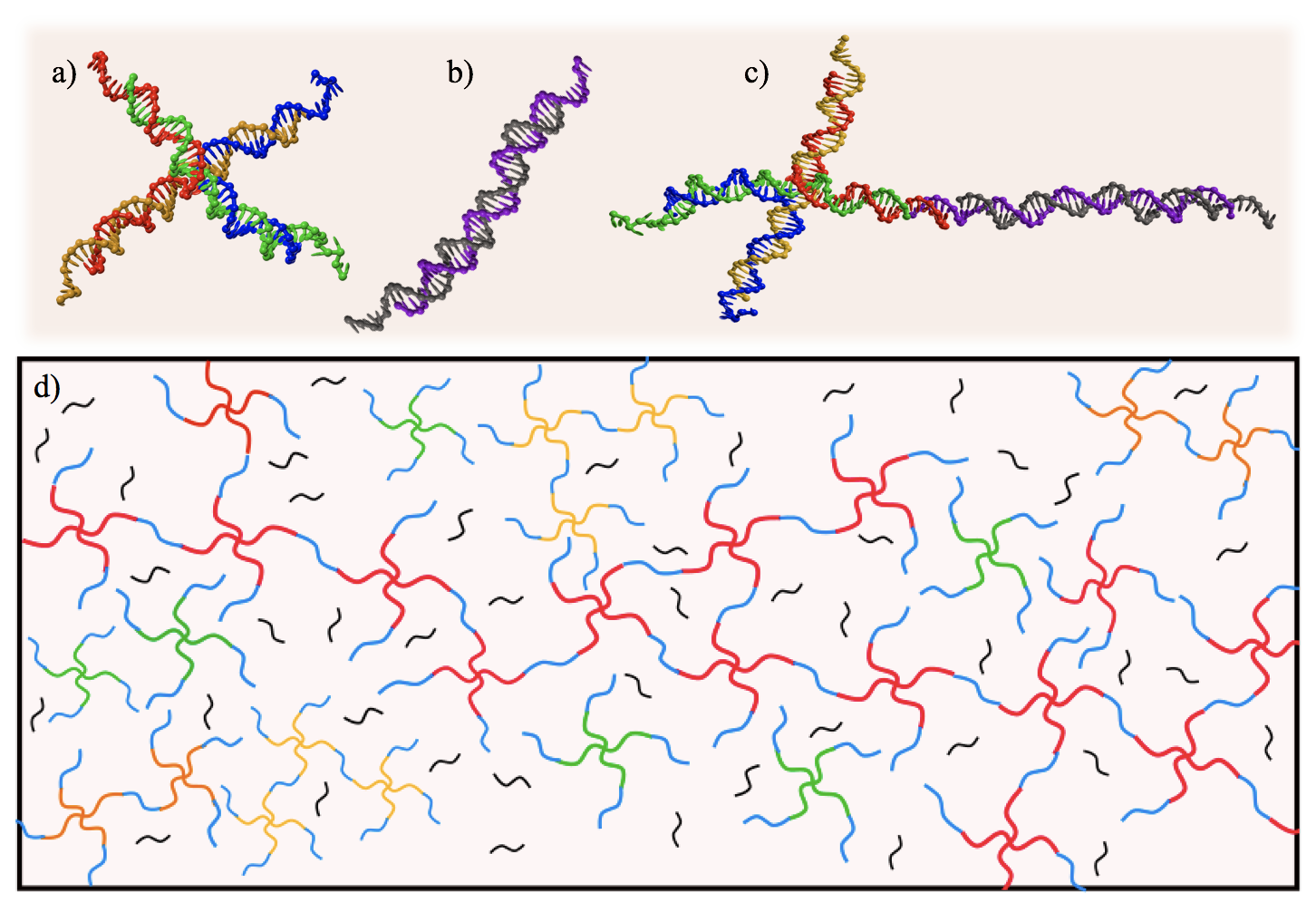} 
  \caption{ Representation of the DNA-made particles, highlighting how the different single strands bind  to form  (a)  the tetra $A$ and (b) the bifunctional $B$  particles. In the (c) pane the $A$ and $B$ particles are shown bonded {\it via} the sticky-end sequence.
   (d) Schematic representation of the  system beyond the percolation threshold ($r<6.0$). The $A$-particles in the spanning   percolating cluster are colored in red,  clusters
   with one $A$ particle  in green,  clusters with two $A$ particles in orange and  clusters with three $A$ particles in golden yellow. The $B$ particles linked to tetramers are plotted in blue while the  non-bonded $B$ particles are represented in black.}
   \label{fig:sketch} 
\end{figure}

We investigate, via dynamic light scattering (DLS),  a binary mixture of valence four ($A$) and valence two ($B$) DNA particles, with selective $AB$ bonds for different ratios $r \equiv N_{B}/N_{A}$, where $N_{A} (N_{B})$ indicates the number of $A$ ($B$) particles (see Fig.~\ref{fig:sketch}-(a-b)). 
Specifically,  we demonstrate, guided by the Flory-Stockmayer polymerization models~\cite{flory1941molecular,stockmayer1943theory},
how either liquid or gel states can be produced by  tuning  $r$.  Indeed,  for the stoichiometric composition $r=2$ (in which there are two $B$  for each $A$), every   $B$  acts as a bridge between  $A$s  and, at low $T$ when all possible bonds are formed, the system gels in a tetravalent structure.   On increasing the concentration of $B$s, the network starts to break. At this point,
some of the  $B$s are connected to only one $A$ particle, preventing the bridging between the tetravalent nodes of the network. 
For a critical $r$, the system reaches the percolation point where a highly polydisperse  cluster size distribution is present. On further increasing  the number of $B$ respect to that of $A$, the system becomes composed by isolated finite size clusters.  DLS provides a clear quantification of these structural changes. 

Different from covalently bonded systems we are able to control, by changing $T$, the $AB$ bond lifetime, being thus able to observe the cross-over  between physical and chemical gels.    

\smallskip{}

\section{Materials and Methods}

\subsection*{Materials and Sample preparation}

\begin{figure}[ht] 
   \centering
   \includegraphics[width=3.0in, angle =0]{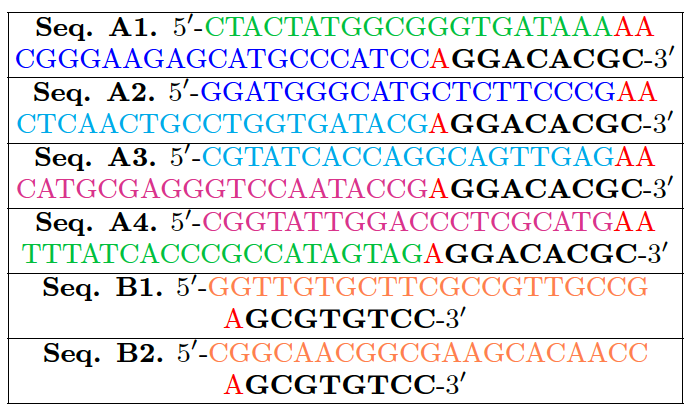} 
   \caption{Sequences of DNA used in this study.}
   \label{table:dna} 
\end{figure}




Tetravalent $A$ nanostars are assembled by mixing in solution  equimolar quantities of 
four different DNA single strands (ss-DNA)  (Seq. A1-A4, in Fig.~\ref{table:dna}) while bivalent $B$  particles
 by  two  ss-DNA (Seq. B1-B2).  As shown in Fig.~\ref{fig:sketch} (a) the $A$ nanostars are an artificial replication of the 4-way DNA intermediates or Holliday junctions~\cite{liu2004happy,wang2016holliday} which play a key role in homologous recombination and DNA repair and which have been of particular interest for biological studies~\cite{van1999assembly, sharples1999holliday}. Differently from the biological relevant cases, the base sequences departing from the junction are asymmetric, preventing the sliding of the junction and thus locking the strands in the desired position.   
Following previous studies~\cite{seeman2003crossroads,seeman2003dna,li2004controlled,biffi2013phase},  the designed sequences incorporate a structural component which determines the particle formation  and
a binding component (the sticky-end sequence) designed to allow interparticle linking, but only between $A$ and $B$. 
Indeed, each arm of the nanostar is composed by two complementary sequences, marked with the same color, of $20$ nucleotides, while
$B$ particles are constituted by a double strand (ds-DNA) composed of $21$-bases. Since the persistence length of 
ds-DNA is about $150$ nucleotides, the $B$ particle can be considered as a rigid rod. 
The two adenine nucleotides in the $A1-A4$ sequences labeled in red in Fig.~\ref{table:dna}) have no complementary 
partners in other sequences. Their role is to  provide flexibility to the core of the $A$ particle~\cite{biffi2013phase}. 
The melting $T$ ($T_{m}$) at which ss-DNA sequences pair to form ds-DNA
is determined by the number of nucleotides in the bonding sequence.  Since the lengths  of the binding parts of the ssDNA
are comparable,  both $A$ and $B$ particles are formed at approximately 
the same $T$ ($65$\textdegree C). Both $A$ and  $B$ ends terminate with an $8$-bases (emboldened in Fig.~\ref{table:dna}) long sticky sequence preceded by 
an extra adenine nucleotide which permits the sticky-ends
to bend and to rotate, easing the linking between $A$ and $B$ binding sites.
The smaller number of nucleotides of the sticky-ends forces   the $A - B$ bonds  to start forming
well below  $65$\textdegree C,  e.g. after the $A$ and $B$ structures have properly self-assembled.

Fig.~\ref{fig:sketch}(a-c) shows the three-dimensional representation of the two particles
and of their binding.

The presence of the adenines at the center of the sequences  generates 
a significant flexibility of the nanostar arms.  Thus we can not exclude the possibility that both ends of a  bridge particle bind with two different arms of the same nanostar.  This case  constitutes the simplest example of a
intra-cluster bond, a  possibility  which is not included in the 
loop-free  calculations of Flory and Stockmayer~\cite{flory1953principles}.  
This  type of intra-cluster link is expected 
to be more frequent at low concentration, when the probability of encountering distinct bridge particles is decreased and in the limit of small ratio of bridges to nanostars.  Since we  work  at large concentrations (see Table~\ref{table1})  and in the limit of large ratio of bridges to nanostars, the presence of such configuration 
is expected to be rear.  We also note  that such intra-cluster bond transforms   a nanostar  into a bridge, essentially acting as a minor renormalization of the ratio $r$. 

\smallskip{}

DNA was provided by Integrated DNA Technologies (IDT) with PAGE purification. Equimolar concentrations of 
$A$ and $B$ ss-DNA sequences were mixed in deionized and filtered H$_{2}$O and NaCl buffer,
 to prepare initial volumes of solutions of $A$ and $B$ particles, 
at concentrations of 191 $\mu$M and 652 $\mu$M, respectively. Nanodrop~\cite{2010_Nanodrop}
measurements confirmed the lack of contaminants.  The final NaCl concentration was  fixed at
200 mM. A centrifugation process ran at $25$\textdegree C/$4.4$ krpm for $10$ minutes facilitated the complete solution of the 
DNA strands in the buffer.
Subsequently, each sample  was kept  $20$ minutes at $90$\textdegree C before being slowly cooled down back to room $T$ overnight.

\smallskip{}
We prepared $4$ samples at  $r$ = $9.0$, $4.9$, $6.0$ and $2.0$. 
In all samples, the  concentration of the total number of particles has been  fixed at the value $\approx 2.4~10^{20}$ per 
litre. Table~\ref{table1} provides detailed information on the prepared samples.

We performed experiments in the  region of concentrations where the system is homogeneous. It has been recently demonstrated that
particles with small valence do phase separate in  gas-like and liquid-like regions  but only in a very limited
range of concentrations~\cite{bianchi2006phase}.  The phase diagram for the studied mixture has been calculated 
within the Wertheim formalism~\cite{wertheim1984fluids1,wertheim1984fluids2}, confirming that the phase-separation region is limited 
to concentrations lower than the one of the fully bonded network~\cite{smallenburg2013patchy}.
To provide a feeling of the values of the selected concentrations  consider that a fully bonded ($r=2.0$) diamond-like structure
composed by  tetrahedral particles linked by the bifunctional  ones (as silicon and oxygen in silica for example) would require a total
molar concentration of  particles (both $A$ and $B$)  of $\approx 400 \mu$M.  Thus, all investigated samples  do not suffer from phase-separation~\cite{smallenburg2013patchy} being
well inside the equilibrium-gel region~\cite{sciortino2017equilibrium}.

\subsection*{Dynamic Light Scattering}

 DLS  measures the intensity correlation functions $g_2(q;t)$.  Experiments
   were performed at a fixed angle $\theta=90$\textdegree~on a setup composed by a Newport $633$ nm He-Ne laser (source power $17$ mW)
and a Brookhaven Inst. correlator. 
Volumes of $45$ $\mu$l for every sample were held in   borosilicate glass tubes with internal radius of $1.5$ mm and immersed in a water bath 
connected to a thermostat.  Thermalisation intervals between $30$ and $40$ minutes were considered to optimise the proper stabilisation of the samples at every $T$, followed by measurements of $25$-$30$ minutes.  The electric field correlation function $g_1(q;t)$ can be 
estimated  via the Siegert relation~\cite{berne2000dynamic}

\begin{equation}
g_{2}(q;t)=1+\beta |g_{1}(q;t)|^{2}, 
\label{eq:Siegert}
\end{equation}

\smallskip{}
where $\beta$ represents the experimental coherence factor.

\begin{table}[htp]
\caption{  Relative ratios $r  \equiv N_B/N_A$, composition $x \equiv N_A/(N_A+N_B)$, and corresponding concentrations $\rho_A$  and $\rho_B$ for  each sample. The last three
columns report mean-field theoretical predictions~\cite{stockmayer1952molecular} for the  case
when all possible bonds are formed (low $T$): (i) 
the fraction of bonded $B$ interacting sites $p_B$; (ii) the mean number of $A$ particles in finite size clusters (MCS) and the
fraction of $A$ particles in the infinite cluster $P_\infty$.}
\begin{center}
\begin{tabular}{|c|c|c|c|c|c|c|}
\hline
  $r$ & $x$ & $\rho_A$ ($\mu$M)  & $\rho_B$ ($\mu$M) & $p_B$ & MCS &$P_\infty$\\
\hline
 9.0 & 0.10 & 39.76 &  357.86 & 0.22 & 3.7 & 0 \\
 6.0 & 0.14 &56.46 &  341.16  & 0.33 &   77  &0 \\
 4.9 & 0.17 & 67.59 &  330.03  & 0.40 &  5.9 & 0.59 \\
 2.0 & 0.33 & 132.54 &  265.08 & 1.0 & 0 & 1\\
 \hline

 \end{tabular}
\end{center}
\label{table1}
\end{table}%

\subsection*{DNA melting curves }

\smallskip{}

\begin{figure}[htbp] 
   \centering
   \includegraphics[width=3.5in]{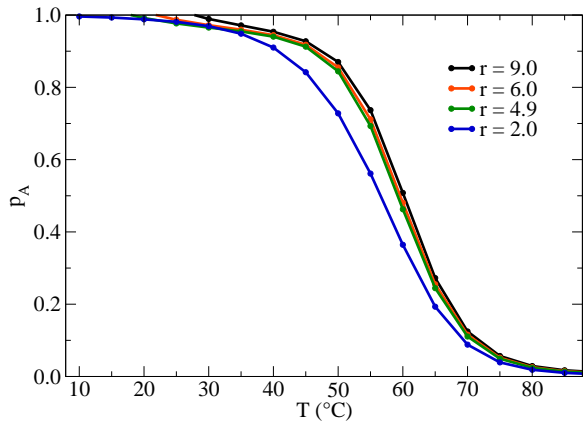} 
   \caption{Probability of observing a  $A$ binding site involved in a $AB$ bond as a function of $T$, evaluated according to the SantaLucia thermodynamic formulation~\cite{SantaLucia17021998} for different ratios $r$}.
   \label{fig:pa} 
\end{figure}

An additional gain of working with DNA nanoparticles is  the well characterized
thermodynamic binding behavior of DNA sequences~\cite{SantaLucia17021998}.
This allows us to properly select the binding temperature of the sticky sequence, controlling the $AB$ bonds.
The $T$ dependence of the bonded fraction of $A$ and $B$ sticky-ends  
$p_A$ and $p_B$  can be quite accurately estimated exploiting  the Santalucia thermodynamic formulation~\cite{SantaLucia17021998} as
coded for example in the open access NUPACK~\cite{nupack} oligo-calculator.  Using as input to NUPACK the concentration of $A$ and $B$ sticky sequences
and the 200 mM NaCl salt concentration, the program provides the $T$-dependence of the fraction of unbonded pairs  $f_{unb}$ which
easily transforms to $p_A=(1- f_{unb})(0.5+r)$ and  $p_B= 2 p_A/r $ (see Sec.~\ref{Sec:model}). Fig.~\ref{fig:pa} shows 
the predicted $T$-dependence of $p_A$ for the investigated samples.
$p_A$ changes from 0 to 1 in a small $T$ interval, centred around $60$\textdegree C, with a width of about $20$\textdegree C. There is a weak dependence on $r$ originating from the progressive larger amount
of $B$ particles on increasing $r$.
Below $T=40$\textdegree C, $p_A$ has reached its asymptotic unit value indicating that at this $T$  essentially all possible bonds in the system are formed. 

\section{Theoretical background: the  $A_{4}-B_{2}$ mixtures with only $AB$ bonds} \label{Sec:model}

To grasp the motivation behind the selected values of $r$, we briefly review 
the seminal work  by Flory~\cite{flory1941molecular} and  Stockmayer~\cite{stockmayer1943theory} on the condensation of  
polyfunctional molecules as a function of  the extent of reaction. In order to better link this theoretical formalism with our experimental research it is desirable to introduce the {\it composition} $x$  of the system as the fraction of $A$ particles present in the system, $x \equiv N_A/(N_A+N_B)$. The values of $x$ are listed in Table~\ref{table1} and are related to the ratio according to $r = (1-x)/x$.  
Neglecting the formation of close loops of bonds (what today we identify with the Bethe lattice~\cite{huang2009introduction})  Flory and Stockmayer were able to
provide mean-field predictions for the location of the percolation (gelation) transition and for the
evolution of the cluster size distribution.  During the years, they focused on several chemical reaction types,
including mixtures of particles of different type and functionality~\cite{flory1953principles}.  An interesting case is offered by
the binary mixture of particles with  functionality $f$ ($f=4$ in the present study) and particles with 
functionality two when only $AB$ bonds are present.   When this is the case, 
the composition of the system is  defined by the  number of particles of each type $N_A$ and $N_B$
and by the probabilities that 
a randomly selected $A$ or $B$ binding site is involved in a bond $p_{A}$  and $p_{B}$.   Since each bond involves  one $A$ and one $B$
bonding site, $p_{A}$ and $p_{B}$ can be calculated as
\begin{equation}
p_{A}=\frac{N_{bonds}}{4N_{A}}~~~~~~p_{B}=\frac{N_{bonds}}{2N_{B}}
\label{eq:p_A}
\end{equation}
where  $N_{bonds}$  indicates the total
number of formed bonds. 
Thus, the presence of only $AB$ bonds  fixes  a connection  between $p_A$ and $p_B$ through   the total number of bonds

\begin{equation}
p_{B}=\frac{2p_{A}N_{A}}{N_{B}} = \frac{2p_{A}}{r}
\label{eq:pb}
\end{equation}
The fully bonded network condition (when all bonding sites are involved in bonds) requires both 
$p_A=1$ and $p_B=1$ and hence $r_{fb}=2.0$. 
Another important relative composition is provided by the percolation threshold, when an infinite spanning cluster first forms.
As shown in Ref.~\cite{stockmayer1952molecular}, this specific point  is  
\begin{equation}
p_{A}p_{B}=\frac{1}{(f-1)}. 
\label{eq:papbperc}
\end{equation}

An interesting case is the one for which the reaction has proceeded to completion, e.g. when all possible bonds are formed.  This is the infinite time limit  
in poly-functional condensation.   When $r=r_{fb}$, this corresponds to a  perfect network, in which all possible binding sites are engaged into bonds. 
For larger values of $r$,  all $A$ binding sites remain bonded while some of the $B$  sites do not participate into bonds. In all these cases ($r>r_{fb}$), $p_A=1$ and
thus $p_B=2/r$.  Similarly, when $r<r_{fb}$  all $B$  sites are bonded, $p_B=1$ and $p_A=r/2$.   
When all possible bonds are formed, 
percolation is thus achieved 
at $r=2/3  $ and $r=6$. For intermediate values of $r$, an infinite spanning cluster is present whereas for values of $r$ outside this interval,
the system is composed by polydisperse clusters of finite size.

The cluster size distribution $N(n,l)$, according to Stockmayer~\cite{stockmayer1952molecular}  is, indicating
with $n$ the number of particles in the cluster of type $A$ and with $l$ the number of particles in the cluster of type $B$,

\begin{eqnarray*}
N(n,l)=  
f N_A \frac{(1-p_A)(1-p_B)}{p_B}   \\ 
 \frac{(f n -n)! (2l -l)! }{ (f n - n -l +1)! (2 l -l -n +1)! n! l!   } x_f^nx_2^l
\label{eq:csdfs}
\end{eqnarray*}

where

\begin{eqnarray*}
x_f= p_B \frac{ (1-p_A)^{(f-1)}}{(1-p_B)} \\
x_2= p_A \frac{(1-p_B)}{(1-p_A)}
\end{eqnarray*}

Note that $ n-1 \le l \le 1+2 n$.  In addition, when $p_A=1$ the value of $l$ is slaved to the value of $n$ by the relation $l=3n+1$.   From Eq.~\ref{eq:csdfs} it is possible
to evaluate the mean cluster size MCS and the fraction of particles in the infinite cluster  $P_\infty$ (the gel). These data for the studied systems are also reported in Table~\ref{table1}. 

The mean field predictions presented above, based on the absence of loops of bonds in finite size clusters, 
are bound to fail close to the percolation threshold where the cluster distribution function must deviate from the
mean field predictions to acquire the  scaling behavior characteristic of the percolation universality class~\cite{rubinstein2003polymer}.
How close the system must be to percolation to feel the non-mean field effects (the Ginzburg criterium~\cite{rubinstein2003polymer}) is a system dependent property.
It has been speculated that in the case of low valence colloidal particles, the mean field predictions are rather robust~\cite{russo2009reversible}.   
In the Appendix we report comparisons between Eq.~\ref{eq:csdfs} and Monte Carlo  simulations at the same $r$ values experimentally studied
to provide evidence that in this small valence systems, 
the mean field predictions are sufficiently accurate (especially far from percolation)  to guide the
interpretation of the experimental results.
 
\section{Results and discussion}

As previously discussed,  on decreasing $r$, the sample changes from a collection of diffusing clusters to a percolating system and 
finally to a fully bonded network.
The number of bonds in the system, quantified  by $p_A$ and $p_B$,  is $T$ dependent.  Still,
as shown in Fig.~\ref{fig:pa}, $p_A$ changes from 0 to 1 in a small $T$ interval. 
Below $40$\textdegree C,  all changes in the dynamics can be attributed to 
the  opening and closing of the bonds and the associated reshuffling of the clusters.  The lifetime of the bonds is itself $T$ dependence and it follows an
Arrhenius law with an activation enthalpy  proportional to the number of bases in the sticky sequence~\cite{vsponer1996structures,schumakovitch2002temperature}.

\begin{figure}[h]
\hspace*{-0.5cm}  
\begin{centering}
\includegraphics[scale=0.34]{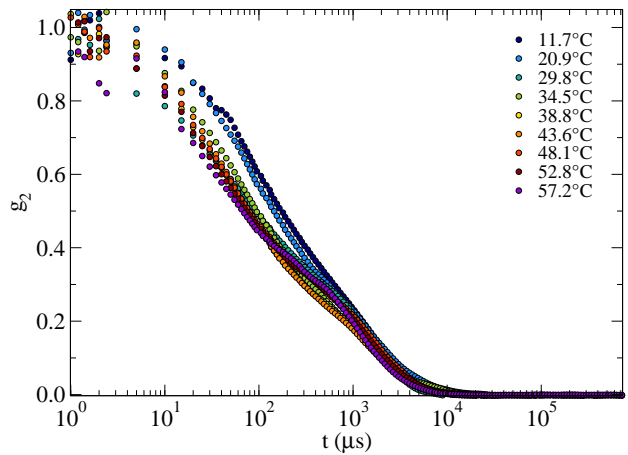} 
\par\end{centering}
\caption{\label{fig:f4010} DLS measurements at $r=9.0$  where a liquid-like behavior is
observed for every $T$. }
\end{figure}

Fig.~\ref{fig:f4010} shows the "fluid" case  $r=9.0$. For this $r$ value, 
even when all bonds are formed,  the large amount of $B$ particles prevents the formation of large clusters.  
The cluster size distribution (Eq.~\ref{eq:csdfs})  predicts that essentially most $A$ particles are involved in 
small aggregates,  coexisting with several isolated $B$ particles.    The MCS (see Table~\ref{table1}) indicates that
the average cluster is composed by about three $A$ particles. Indeed, we observe
 that   the correlation function  does not significantly vary with $T$ and it approaches zero  at about $10^4 \mu$s, consistent with the
 expectation that the clusters in the system are of limited size and do not grow on cooling. 
 
\begin{figure}[h] 
\hspace*{-0.5cm}  
\begin{centering}
\includegraphics[scale=0.35]{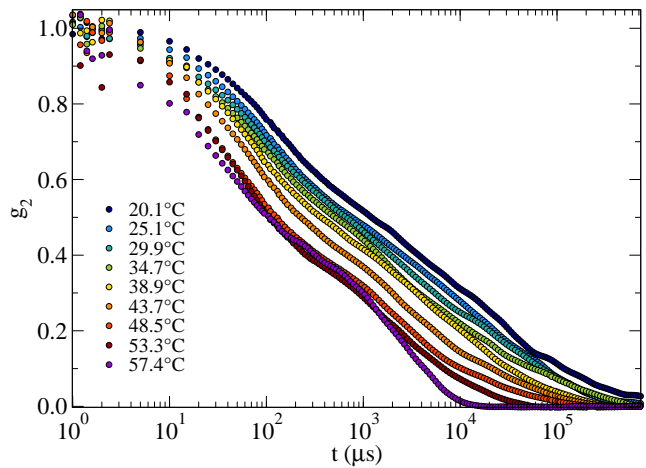} 
\par\end{centering}
\caption{\label{fig:f4014} DLS measurements at  $r=6.0$. Note that 
 below $50$\textdegree C all curves follow a logarithmic decay.}
\end{figure}

 The  sample   at $r= 6.0$ corresponds  to six $B$ particles for each $A$ particle.
This $r$ value has been selected since,  according to  mean-field predictions  (Eq.~\ref{eq:papbperc}),
 when all possible bonds are formed, the system is close to percolation and thus composed by highly polydisperse
 clusters, power-law distributed in size.  Different from cases in which percolation results as an effect of an irreversible
 polymerization process  (and as such  controlled by curing time), here we tune the ratio $r$ between $B$ and $A$ particles 
 to approach the percolation point. At low $T$, when the lifetime of the bonds is larger than the experimental time,
 the system is expected to behave as a chemical gel at percolation~\cite{henningawinter1995rheological, russo2009reversible, lindquist2016formation}. 
The corresponding DLS results are  presented in Fig. \ref{fig:f4014}. At the highest $T$  ($T>53$\textdegree C), the density fluctuations
still decay at times comparable to those observed in the sample at $r = 9.0$ previously discussed. However, when  $T<50$\textdegree C
the initial decay (which is possibly associated to the excess free $B$ particles and to the vibrational motion within clusters),
is followed by a clear  logarithmic decay which extends for several decades, a signature of a highly polydisperse system.

\begin{figure}[h]
\hspace*{-0.5cm}  
\vspace*{0cm}  
\begin{centering}
\includegraphics[scale=0.34]{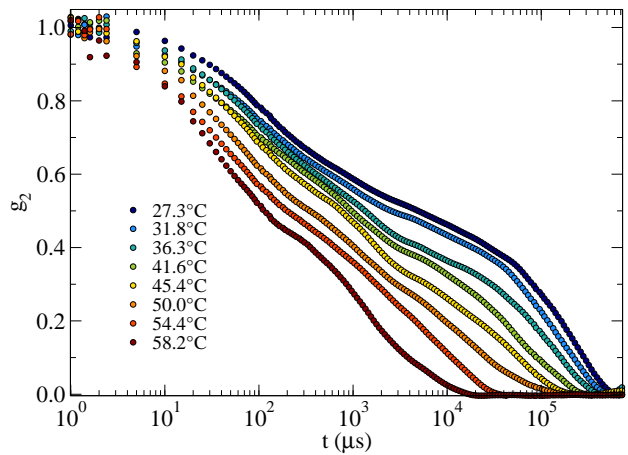} 
\par\end{centering}
\caption{\label{fig:f4017} DLS measurements at  $r=4.9$ where the system is beyond  percolation.}
\end{figure}

\begin{figure}[h]
\hspace*{-0.5cm}
\vspace*{-0cm}    
\begin{centering}
\includegraphics[scale=0.34]{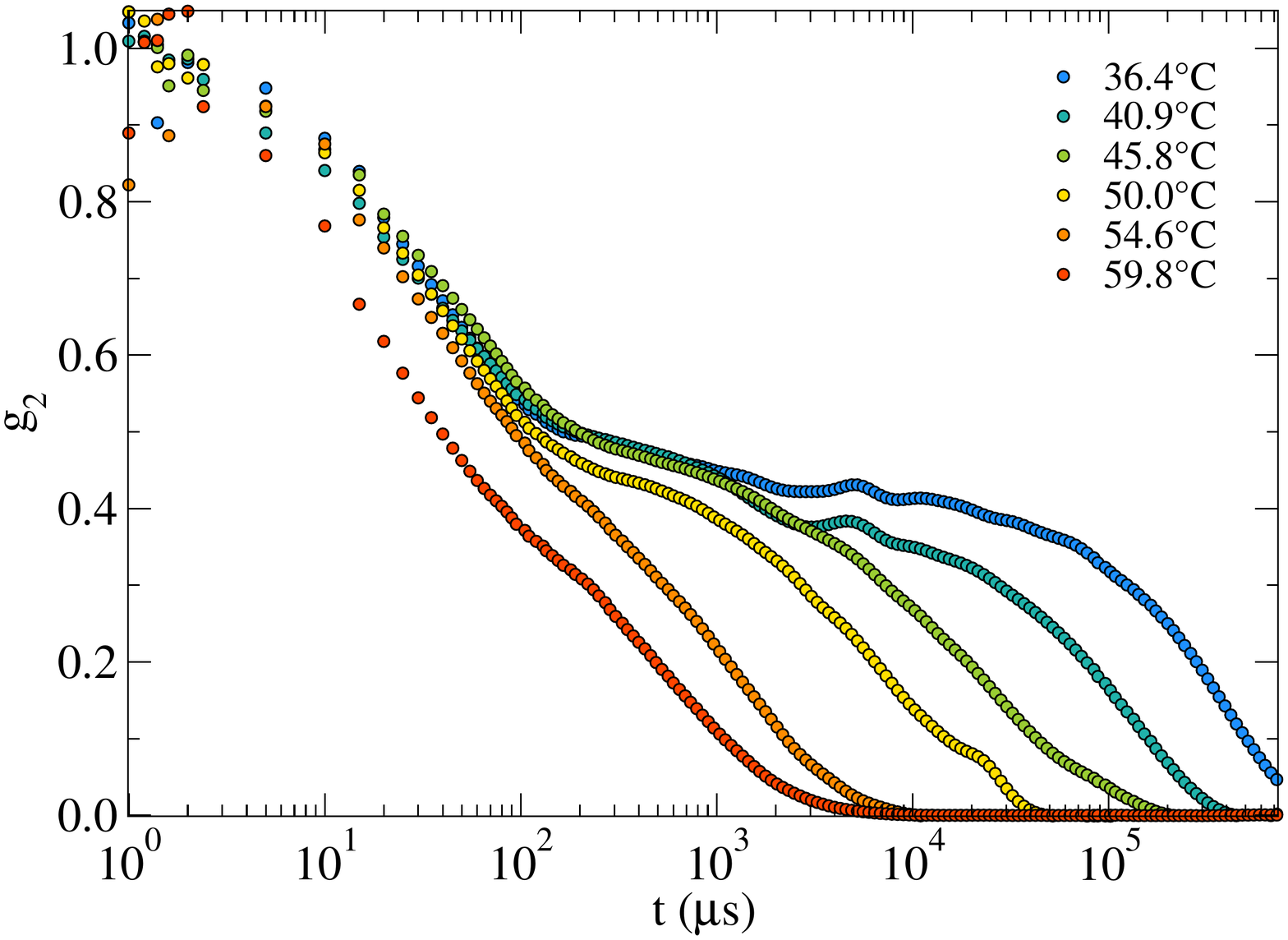} 
\par\end{centering}
\setlength{\belowcaptionskip}{43pt}
\caption{\label{fig:f4033}   DLS measurements  at $r=2.0$  where
the system continuously evolves from a liquid-like behavior at high $T$ to a gel at lower $T$.
The  relaxation time  increases 
 more than $3$
orders of magnitude, while the   plateau  increases its height. }
\end{figure}

DLS techniques have been  successfully applied to investigate gelation in chemical and physical gels~\cite{adam1988dynamical,
richter2007comparison,berthon2001dls,djabourov1993structure,bordi2002chemical}.
 In the limit of very dilute samples, when correlation between different clusters can be neglected, the DLS correlation function can be
interpreted as resulting by the diffusive motion of the clusters (e.g. the self-component of the collective behavior). Under these conditions,
the self-similarity  in the cluster size, coupled to the scaling behavior of the cluster diffusion coefficient  and of the size dependence  of the
cluster scattering intensity generates a very wide (often power-law) decay of the density fluctuations~\cite{adam1988dynamical}. When cluster interactions are
present the correlation function retains its very wide spectrum of decay times which can manifest as a power-law, a logarithmic behavior or 
a stretched exponential decay with small stretching exponent~\cite{sciortino1997relaxation}. 
Our sample has a non-negligible density  and cluster-cluster excluded volume interactions 
prevent the possibility of describing the system as an ideal gas of polydisperse clusters and  as such the decay of the correlation function can not be written as a sum of a 
independent decays. Still, the large polydispersity characterizing the proximity of the percolation point reveals itself in the observed logarithmic decay.

\begin{figure}
\hspace*{-0.5cm}
\vspace*{-0cm}    
\begin{centering}
\includegraphics[scale=0.34]{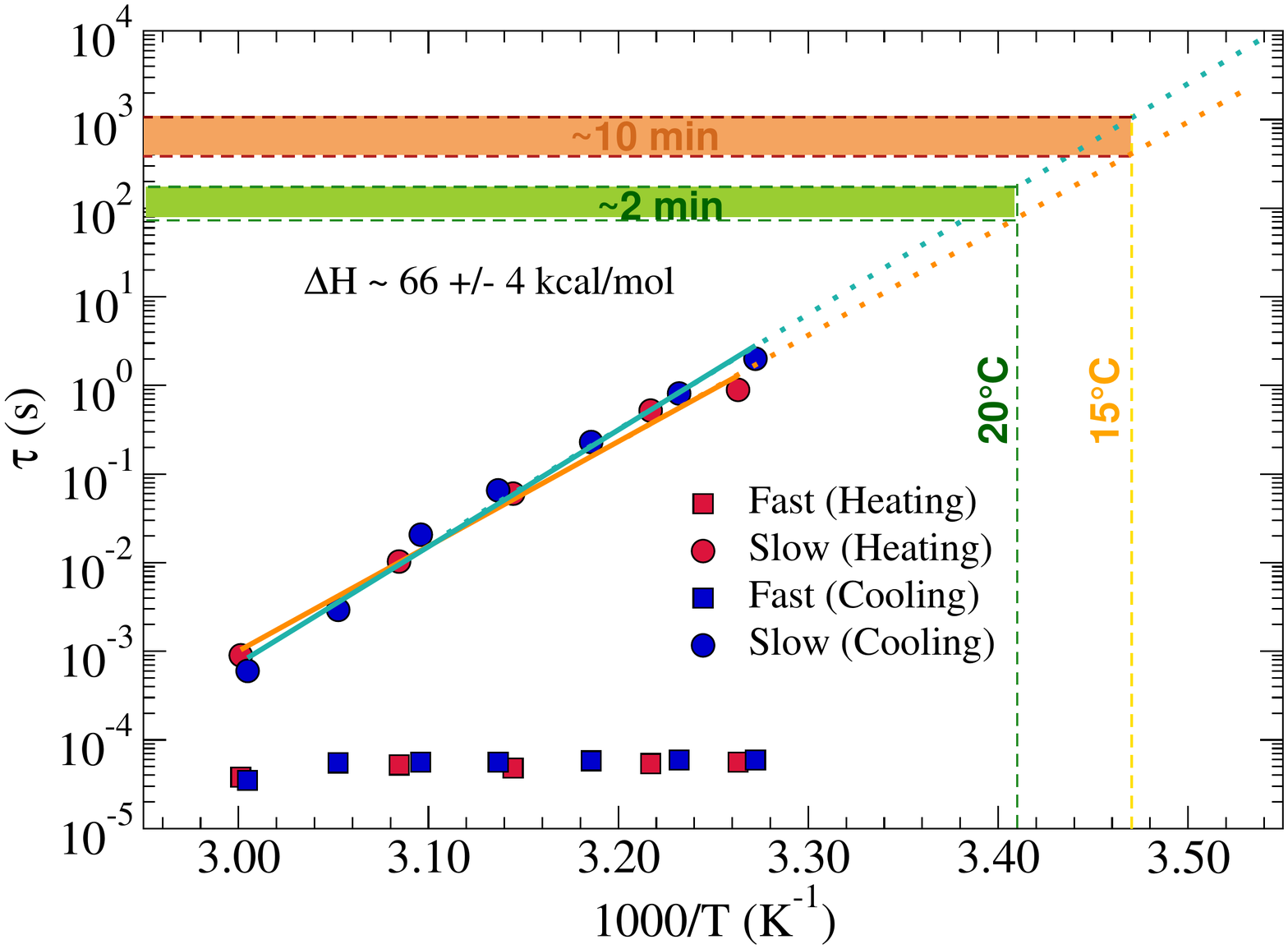} 
\par\end{centering}
\setlength{\belowcaptionskip}{43pt}
\caption{\label{fig:arrhenius} Decay times  $\tau_f$ and $\tau_s$  as a function of $T$ of the $x_{fb}$ system. Results on heating (red) and on cooling (blue) are reported. 
$\tau_s$ (full circles) follows an Arrhenius behavior  while $\tau_f$ 
 (full squares) does not show any detectable $T$ dependence. Arrhenius fit to $\tau_s$   are indicated with dashed lines.  The green and yellow shaded surfaces
indicates the $2$ and $10$ minutes timescales, corresponding to  $T=20$\textdegree C and 15\textdegree C.}
\end{figure}

\smallskip{}

The third sample, prepared with  $r = 4.9$, is  beyond the percolation threshold.  
The system is composed by a spanning cluster complemented by a polydisperse set of finite size clusters, composing the
sol fraction.  According to Eq.~\ref{eq:csdfs}, at low $T$ the average number of $A$ particles in finite size clusters is about six (see Table~\ref{table1}),
while the fraction of $A$ particles in the infinite cluster is about $59 \%$.
The corresponding measured  correlation functions  are displayed in Fig.~\ref{fig:f4017}. Again, at $T$ > $50$\textdegree C, 
most of the particles are isolated, or forming unconnected small clusters, and so the density fluctuations decorrelate at times 
comparable to those measured at the same $T$ in all  studied samples. 
The mixed nature of the sample (sol+gel) 
determines a complex shape of the  correlation function, in which the logarithmic decay, which we attribute to the diffusion of the
finite size clusters, is complemented by a more exponential decay originating from the gel component. 

\smallskip{}

Finally, Fig.~\ref{fig:f4033} shows the case for  $r=2.0$, matching the stoichiometry of the system with the particles valences.
For this $r$, at low $T$ all particles belong to the infinite cluster. The decay of the correlation function closely reproduces the one previously measured  in a  one-component valence-four DNA nanostar system in which  the sticky-end sequence  was selected to be self-complementary,   enabling bonding among identical particles~\cite{biffi2013phase}.  
In the present experiment  the bond between $A$ particles is mediated by the presence of the $B$ particles. Despite this difference, the
decay of the correlation function is very similar, displaying a two-step relaxation process  (a fast and a slow process) in which the 
slow decay time progressively increases on cooling. We first note that experiments in this sample are limited to $T>36$\textdegree C.  For smaller $T$,
the characteristic time becomes longer than a few seconds, which is the maximum time accessible with our light scattering apparatus.
This indicates that for $T< 36$\textdegree C we can consider the system as non-restructuring for the duration of the measurement (chemical gel).
As in Ref.~\onlinecite{biffi2013phase}, the correlation functions can be fitted as 

\begin{equation}
g_{1}(\tau) = (1-\alpha)\, \exp \left [ -\left(\frac{\tau}{\tau _{f}}\right)\right ] + \alpha \, \exp \left [ -\left(\frac{\tau}{\tau _{s}}\right)^{\beta _{s}}\right ], 
\label{eq:stretch}
\end{equation}
 where $\alpha$ is the amplitude of the slow process and ${\tau _{f}}$, ${\tau _{s}}$ and ${\beta _{s}}$ stand for the fast and slow relaxation 
 times and for the stretching exponential factor, respectively. 
 In agreement with the valence-four nanostar  case~\cite{biffi2013phase,bomboi2016re}
 ${\beta _{s}}$ is comprised in the range $0.63-0.79$. 
 Also in agreement with previous data,   the fast relaxation mode  does not show any significant dependence on $T$,
while ${\tau _{s}}$  follows an Arrhenius  dependence (Fig. \ref{fig:arrhenius}).
The associated activation energy, $\approx 66 \pm 5$ kcal/mol, has a value comparable with the Gibbs free energy associated to the breaking/formation of the $AB$ bonds ($69$ kcal/mol~\cite{SantaLucia17021998}).   For our purposes it is important to note that  the slowing down of the dynamics  reflects the progressive 
increase in the bond lifetime. Indeed, the bond breaking and reforming processes constitute the elementary steps by which the local topology of the network
changes, allowing the relaxation of the density fluctuations~\cite{nava2017fluctuating}.

\smallskip{}

\begin{figure}[t]
\hspace*{-0.5cm}
\vspace*{-0.5cm}    
\begin{centering}
\includegraphics[scale=0.34]{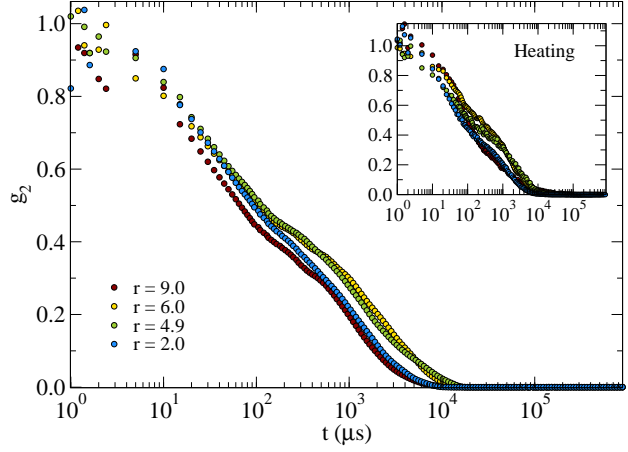} 
\par\end{centering}
\caption{\label{fig:T58} DLS measurements  at $T=58$\textdegree C. The  decay of the correlation  
 is rather insensitive to $r$. Measurements at $T=58$\textdegree C after heating from $10$\textdegree C are displayed in the inset.}
\end{figure}

\begin{figure}[t]
\hspace*{-0.5cm}
\vspace*{-0.5cm}    
\begin{centering}
\includegraphics[scale=0.34]{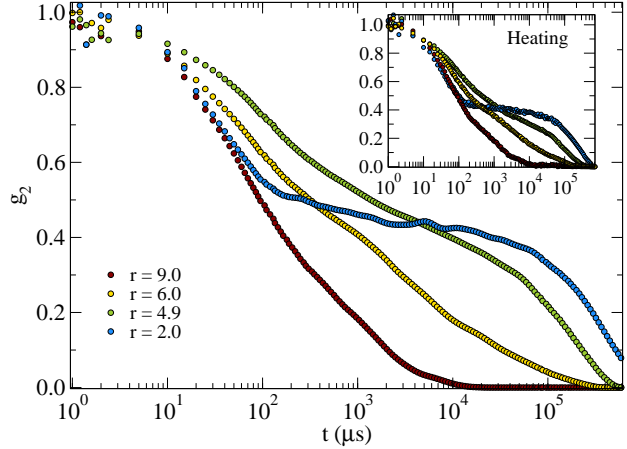} 
\par\end{centering}
\caption{\label{fig:T35}   DLS measurements  at $T=35$\textdegree C. At this $T$ all  possible $AB$ bonds are formed.
Measurements at   $35$\textdegree C  after heating the samples up from $10$\textdegree C are displayed in the inset.  }
\end{figure}

Next we compare the decay of the correlation functions at different $r$  but at the same $T$, to attempt to disentangle the role of the
network connectivity from the role of the bond lifetime.  In  Fig.~\ref{fig:T58} and Fig.~\ref{fig:T35} we show two different $T$ values,
one corresponding to a high $T$ ($58$\textdegree C) where only a fraction of the possible bonds are formed (see Fig.~\ref{fig:pa}) and one 
corresponding to  $35$\textdegree C. In the latter case,   the structural evolution of the system is already completed and the only dynamic events altering the clusters connectivity are
related to  breaking and  reforming of  bonds.   The value of $\tau_s$ measured in the
 $r=2.0$ case at this two $T$s, an estimate of  the bond lifetime, are respectively $10^3 \mu$s and  $10^6 \mu$s.   Fig.~\ref{fig:T58}  demonstrates that the decay of the correlation functions at this
$T$ is very similar in all samples.  This is consistent with the idea that, independently of  $r$, the clusters are small and  they 
change their identity due to the fast bond breaking, well within the experimental sampled time window.   
Fig.~\ref{fig:T35} shows the case of $T=35$\textdegree C. Here  extremely diversified behaviors can be observed:
the cross-over from a liquid (sol) to a fully bonded network state, through the percolation  threshold is spelled out.
The fast decay observed for  $r=9.0$, indicating the presence of small clusters, turns into a remarkable logarithmic decay  when $r$ is close to the
percolation transition, changing finally into a two-step relaxation process for  smaller $r$, when a gel component is present.   

\smallskip{}
We have  repeated the experimental study  on the same samples following an inverted thermalisation route, 
 ten   days after the original  
measurements. Starting from low $T$ the samples
were progressively thermalised on heating, up to $57$\textdegree C. The insets of Fig.~\ref{fig:T58} and Fig.~\ref{fig:T35}
show   a satisfactory agreement between cooling and heating 
ramps, demonstrating the equilibrium nature of the systems, supporting that the all-DNA design here presented is completely thermoreversible.

\section{Summary and Conclusions}

We have demonstrated how DNA-made particles can be designed and created in the lab to 
closely behave as tetra- and bi-functional units for which it is possible to 
control  selective binding ($AB$ binding in the present case) as well as  
the $T$ at which the fully reversible binding process takes place. 
Specifically, we select 
four complementary single strands of DNA to self-assemble  tetravalent $A$ nanostars and
two sequences to compose the bivalent $B$ particles. The $B$ particles act as bridges which connect $A$ particles via 
complementary sticky overhangs attached to the particle arms. The investigated system 
provides the (reversible) colloidal analog of the (irreversible) $A_4-B_2$ polyfunctional polymerization condensation~\cite{rubinstein2003polymer}.
 By modulating the ratio of $A$ and $B$ particles we have investigated different states 
on the sol-gel route, from a fluid of small clusters to a fully bonded stoichiometric gel.
In addition, by selecting the  binding temperature between  $40$ and $60$\textdegree C, we have investigated both the effect of progressive bond formation
as well as the almost chemical case, where the bond lifetime becomes longer than the accessible experimental observation time window.

\smallskip{}
Guided by the mean-field predictions of  the classical models of Flory and Stockmayer~\cite{stockmayer1952molecular} (and corroborated by the comparison with simulation data reported in the Appendix), we have 
performed  light scattering experiments, probing the decay of density fluctuations. We have observed on cooling a progressive slowing down of the
dynamics associated to the formation of larger and larger clusters. Below $40$\textdegree C, when all possible bonds are formed, the structural evolution of the system
is completed and the dynamics becomes controlled by the bond lifetime, which is represented by an Arrhenius $T$-dependence with an activation
energy fixed by the number of base pairs of the sticky-end sequence. 
 Around percolation, we observe a very clear logarithmic decay of the
correlation function.  This functional form is observed also in glass-forming systems when different competing arrest mechanisms are present~\cite{fabbian1999ideal,sciortino2003evidence,pham2002multiple,gotze2004nearly}.  Logarithmic decays have also been reported
in mixtures of polymeric systems~\cite{moreno2006anomalous} and in glassy wormlike chain polymers~\cite{glaser2008dynamic}.
Finally, in the fully bonded case we observe a clear two-step relaxation process.  All percolating samples are examples of low valence equilibrium gel-forming systems~\cite{sciortino2017equilibrium}, fully reversible with $T$.

\smallskip{}
In conclusion we observe that the condition of full bonding is 
achieved in the present DNA system for $T<40$\textdegree C (Fig.~\ref{fig:pa}).   According to the 
measure activation enthalpy, below $T<15$\textdegree  C, the bond lifetime is longer than ten minutes, effectively transforming the system  in a stable chemical gel. 
Different from the chemical case, in which aggregation is irreversible and stresses are frozen in, 
here the process is fully reversible and a slow annealing allows the system to sample equilibrium  stress-free configurations.
We expect these results to be interesting for further investigations, specifically in the bottom-up
self-assembling of  biocompatible nanomaterials, in DNA-based plastics~\cite{romano2015switching}, and in all cases in which a medium close to percolation is demanded~\cite{gnancasimir2014}.

\section{Appendix}
In this Appendix we report a comparison between a Monte Carlo simulation of
a binary mixture of patchy particles with valence four and two and the Stockmayer 
theoretical predictions (Eq.~\ref{eq:csdfs}) with the aim of accessing the quality of the theoretical
predictions and the possibility to build on the theoretical data to interpret the
experimental results.

\begin{figure}[t]
\hspace*{-0.5cm}
\vspace*{-0.5cm}    
\begin{centering}
\includegraphics[scale=0.34]{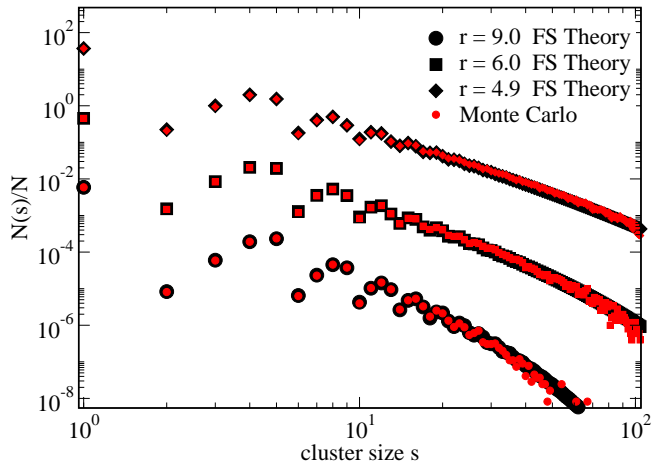} 
\par\end{centering}
\caption{\label{fig:csd}  Cluster size distribution $N(s)/N$ for  $r=9.0, 6.0$ and $4.9$
for $\rho \sigma^3=0.05$ and $k_BT/\epsilon=0.1$ as function of the number of particles in the cluster $s$.  Large black symbols denote 
Flory-Stockmayer (FS)
theoretical predictions
(Eq.~\ref{eq:csdfs}) while smaller red symbols denote the Monte Carlo results.  
The curves for  $r$=9.0 have been divided by 100 while the curves for $r$=4.9 have been 
multiplied by 100 to improve visibility of the comparison.  The non-monotonic behavior 
as small cluster size $s$ originates by the smaller probability of observing 
clusters with un-bonded $A$ binding sites.
  }
\end{figure}

In the simulation the particles are modelled as patchy spheres  of diameter $\sigma$
with patches  arranged in tetrahedral ($A$) and polar ($B$)
geometry. Particles interact via  the widely
used Kern-Frenkel model~\cite{kern2003fluid}. The angular square-well  
is defined by an opening angle  $\cos \theta=0.92$, a width 
$0.15 \sigma$, and a depth $\epsilon$. 
We investigate $N=50000$ particles in a cubic box of size $L=100\sigma$ 
(corresponding to a number density $\rho \sigma^3=0.05$) with 
periodic boundary conditions at  three of the  $r$ values studied in the experiments.
Finally, to sample a reasonable number of distinct configurations, we fix
$k_BT/\epsilon=0.1$, corresponding to $p_A > 0.8$.  
A cluster algorithm is used to identify the clusters of bonded particles
and their distribution. The number of bonds observed in equilibrium in the simulation 
(which coincides rather precisely with  theoretical expectations
calculated using the law of mass action describing a chemical equilibrium between bonded and unbonded sites, see for example Eq.~4 and 5 in
Ref.~\cite{russo2009reversible}) provides via Eq.~\ref{eq:p_A}
 the 
corresponding values of $p_A$ and $p_B$. The resulting theoretical
cluster size distributions (Eq.~\ref{eq:csdfs}) are compared with the corresponding
Monte Carlo results in Fig.~\ref{fig:csd}.

\subsection*{Acknowledgments}

FS and JFC acknowledge support from ETN-COLLDENSE (H2020-MCSA-ITN-2014,
Grant No. 642774).  We thank L. Rovigatti for providing us with the schematic graphs 
of the $A$ and $B$ particles in Fig.~1 and  F. Bordi for discussions and support with the experimental 
setup. We also thank G. Amico for technical support.
\subsection*{Additional Information}
Competing financial interests: the authors declare no competing financial interests.

\bibliography{expA4B2DNA_Bib}

\begin{thebibliography}{63}%
\makeatletter
\providecommand \@ifxundefined [1]{%
 \@ifx{#1\undefined}
}%
\providecommand \@ifnum [1]{%
 \ifnum #1\expandafter \@firstoftwo
 \else \expandafter \@secondoftwo
 \fi
}%
\providecommand \@ifx [1]{%
 \ifx #1\expandafter \@firstoftwo
 \else \expandafter \@secondoftwo
 \fi
}%
\providecommand \natexlab [1]{#1}%
\providecommand \enquote  [1]{``#1''}%
\providecommand \bibnamefont  [1]{#1}%
\providecommand \bibfnamefont [1]{#1}%
\providecommand \citenamefont [1]{#1}%
\providecommand \href@noop [0]{\@secondoftwo}%
\providecommand \href [0]{\begingroup \@sanitize@url \@href}%
\providecommand \@href[1]{\@@startlink{#1}\@@href}%
\providecommand \@@href[1]{\endgroup#1\@@endlink}%
\providecommand \@sanitize@url [0]{\catcode `\\12\catcode `\$12\catcode
  `\&12\catcode `\#12\catcode `\^12\catcode `\_12\catcode `\%12\relax}%
\providecommand \@@startlink[1]{}%
\providecommand \@@endlink[0]{}%
\providecommand \url  [0]{\begingroup\@sanitize@url \@url }%
\providecommand \@url [1]{\endgroup\@href {#1}{\urlprefix }}%
\providecommand \urlprefix  [0]{URL }%
\providecommand \Eprint [0]{\href }%
\providecommand \doibase [0]{http://dx.doi.org/}%
\providecommand \selectlanguage [0]{\@gobble}%
\providecommand \bibinfo  [0]{\@secondoftwo}%
\providecommand \bibfield  [0]{\@secondoftwo}%
\providecommand \translation [1]{[#1]}%
\providecommand \BibitemOpen [0]{}%
\providecommand \bibitemStop [0]{}%
\providecommand \bibitemNoStop [0]{.\EOS\space}%
\providecommand \EOS [0]{\spacefactor3000\relax}%
\providecommand \BibitemShut  [1]{\csname bibitem#1\endcsname}%
\let\auto@bib@innerbib\@empty
\bibitem [{\citenamefont {Seeman}(1998)}]{seeman1998dna}%
  \BibitemOpen
  \bibfield  {author} {\bibinfo {author} {\bibfnamefont {N.~C.}\ \bibnamefont
  {Seeman}},\ }\href@noop {} {\bibfield  {journal} {\bibinfo  {journal} {Annu.
  Rev. Bioph. Biom.}\ }\textbf {\bibinfo {volume} {27}},\ \bibinfo {pages}
  {225} (\bibinfo {year} {1998})}\BibitemShut {NoStop}%
\bibitem [{\citenamefont {Bashir}(2001)}]{bashir2001invited}%
  \BibitemOpen
  \bibfield  {author} {\bibinfo {author} {\bibfnamefont {R.}~\bibnamefont
  {Bashir}},\ }\href@noop {} {\bibfield  {journal} {\bibinfo  {journal}
  {Superlattice. Microst.}\ }\textbf {\bibinfo {volume} {29}},\ \bibinfo
  {pages} {1} (\bibinfo {year} {2001})}\BibitemShut {NoStop}%
\bibitem [{\citenamefont {Pinheiro}\ \emph {et~al.}(2011)\citenamefont
  {Pinheiro}, \citenamefont {Han}, \citenamefont {Shih},\ and\ \citenamefont
  {Yan}}]{pinheiro2011challenges}%
  \BibitemOpen
  \bibfield  {author} {\bibinfo {author} {\bibfnamefont {A.~V.}\ \bibnamefont
  {Pinheiro}}, \bibinfo {author} {\bibfnamefont {D.}~\bibnamefont {Han}},
  \bibinfo {author} {\bibfnamefont {W.~M.}\ \bibnamefont {Shih}}, \ and\
  \bibinfo {author} {\bibfnamefont {H.}~\bibnamefont {Yan}},\ }\href@noop {}
  {\bibfield  {journal} {\bibinfo  {journal} {Nat. Nanotechnol.}\ }\textbf
  {\bibinfo {volume} {6}},\ \bibinfo {pages} {763} (\bibinfo {year}
  {2011})}\BibitemShut {NoStop}%
\bibitem [{\citenamefont {Biffi}\ \emph {et~al.}(2013)\citenamefont {Biffi},
  \citenamefont {Cerbino}, \citenamefont {Bomboi}, \citenamefont {Paraboschi},
  \citenamefont {Asselta}, \citenamefont {Sciortino},\ and\ \citenamefont
  {Bellini}}]{biffi2013phase}%
  \BibitemOpen
  \bibfield  {author} {\bibinfo {author} {\bibfnamefont {S.}~\bibnamefont
  {Biffi}}, \bibinfo {author} {\bibfnamefont {R.}~\bibnamefont {Cerbino}},
  \bibinfo {author} {\bibfnamefont {F.}~\bibnamefont {Bomboi}}, \bibinfo
  {author} {\bibfnamefont {E.~M.}\ \bibnamefont {Paraboschi}}, \bibinfo
  {author} {\bibfnamefont {R.}~\bibnamefont {Asselta}}, \bibinfo {author}
  {\bibfnamefont {F.}~\bibnamefont {Sciortino}}, \ and\ \bibinfo {author}
  {\bibfnamefont {T.}~\bibnamefont {Bellini}},\ }\href@noop {} {\bibfield
  {journal} {\bibinfo  {journal} {Proc. Natl. Acad. Sci. USA}\ }\textbf
  {\bibinfo {volume} {110}},\ \bibinfo {pages} {15633} (\bibinfo {year}
  {2013})}\BibitemShut {NoStop}%
\bibitem [{\citenamefont {Bomboi}\ \emph {et~al.}(2016)\citenamefont {Bomboi},
  \citenamefont {Romano}, \citenamefont {Leo}, \citenamefont
  {Fernandez-Castanon}, \citenamefont {Cerbino}, \citenamefont {Bellini},
  \citenamefont {Bordi}, \citenamefont {Filetici},\ and\ \citenamefont
  {Sciortino}}]{bomboi2016re}%
  \BibitemOpen
  \bibfield  {author} {\bibinfo {author} {\bibfnamefont {F.}~\bibnamefont
  {Bomboi}}, \bibinfo {author} {\bibfnamefont {F.}~\bibnamefont {Romano}},
  \bibinfo {author} {\bibfnamefont {M.}~\bibnamefont {Leo}}, \bibinfo {author}
  {\bibfnamefont {J.}~\bibnamefont {Fernandez-Castanon}}, \bibinfo {author}
  {\bibfnamefont {R.}~\bibnamefont {Cerbino}}, \bibinfo {author} {\bibfnamefont
  {T.}~\bibnamefont {Bellini}}, \bibinfo {author} {\bibfnamefont
  {F.}~\bibnamefont {Bordi}}, \bibinfo {author} {\bibfnamefont
  {P.}~\bibnamefont {Filetici}}, \ and\ \bibinfo {author} {\bibfnamefont
  {F.}~\bibnamefont {Sciortino}},\ }\href@noop {} {\bibfield  {journal}
  {\bibinfo  {journal} {Nat. Comm.}\ }\textbf {\bibinfo {volume} {7}} (\bibinfo
  {year} {2016})}\BibitemShut {NoStop}%
\bibitem [{\citenamefont {Russo}\ \emph {et~al.}(2011)\citenamefont {Russo},
  \citenamefont {Tavares}, \citenamefont {Teixeira}, \citenamefont {Telo~da
  Gama},\ and\ \citenamefont {Sciortino}}]{russo2011reentrant}%
  \BibitemOpen
  \bibfield  {author} {\bibinfo {author} {\bibfnamefont {J.}~\bibnamefont
  {Russo}}, \bibinfo {author} {\bibfnamefont {J.}~\bibnamefont {Tavares}},
  \bibinfo {author} {\bibfnamefont {P.~I.~C.}\ \bibnamefont {Teixeira}},
  \bibinfo {author} {\bibfnamefont {M.~M.}\ \bibnamefont {Telo~da Gama}}, \
  and\ \bibinfo {author} {\bibfnamefont {F.}~\bibnamefont {Sciortino}},\
  }\href@noop {} {\bibfield  {journal} {\bibinfo  {journal} {Phys. Rev. Lett.}\
  }\textbf {\bibinfo {volume} {106}},\ \bibinfo {pages} {085703} (\bibinfo
  {year} {2011})}\BibitemShut {NoStop}%
\bibitem [{\citenamefont {de~Las~Heras}, \citenamefont {Tavares},\ and\
  \citenamefont {Telo~da Gama}(2011)}]{de2011phase}%
  \BibitemOpen
  \bibfield  {author} {\bibinfo {author} {\bibfnamefont {D.}~\bibnamefont
  {de~Las~Heras}}, \bibinfo {author} {\bibfnamefont {J.~M.}\ \bibnamefont
  {Tavares}}, \ and\ \bibinfo {author} {\bibfnamefont {M.~M.}\ \bibnamefont
  {Telo~da Gama}},\ }\href@noop {} {\bibfield  {journal} {\bibinfo  {journal}
  {Soft Matter}\ }\textbf {\bibinfo {volume} {7}},\ \bibinfo {pages} {5615}
  (\bibinfo {year} {2011})}\BibitemShut {NoStop}%
\bibitem [{\citenamefont {de~Las~Heras}, \citenamefont {Tavares},\ and\
  \citenamefont {Telo~da Gama}(2012)}]{de2012bicontinuous}%
  \BibitemOpen
  \bibfield  {author} {\bibinfo {author} {\bibfnamefont {D.}~\bibnamefont
  {de~Las~Heras}}, \bibinfo {author} {\bibfnamefont {J.~M.}\ \bibnamefont
  {Tavares}}, \ and\ \bibinfo {author} {\bibfnamefont {M.~M.}\ \bibnamefont
  {Telo~da Gama}},\ }\href@noop {} {\bibfield  {journal} {\bibinfo  {journal}
  {Soft Matter}\ }\textbf {\bibinfo {volume} {8}},\ \bibinfo {pages} {1785}
  (\bibinfo {year} {2012})}\BibitemShut {NoStop}%
\bibitem [{\citenamefont {Zhang}\ \emph {et~al.}(2005)\citenamefont {Zhang},
  \citenamefont {Keys}, \citenamefont {Chen},\ and\ \citenamefont
  {Glotzer}}]{zhang2005self}%
  \BibitemOpen
  \bibfield  {author} {\bibinfo {author} {\bibfnamefont {Z.}~\bibnamefont
  {Zhang}}, \bibinfo {author} {\bibfnamefont {A.~S.}\ \bibnamefont {Keys}},
  \bibinfo {author} {\bibfnamefont {T.}~\bibnamefont {Chen}}, \ and\ \bibinfo
  {author} {\bibfnamefont {S.~C.}\ \bibnamefont {Glotzer}},\ }\href@noop {}
  {\bibfield  {journal} {\bibinfo  {journal} {Langmuir}\ }\textbf {\bibinfo
  {volume} {21}},\ \bibinfo {pages} {11547} (\bibinfo {year}
  {2005})}\BibitemShut {NoStop}%
\bibitem [{\citenamefont {Smallenburg}\ and\ \citenamefont
  {Sciortino}(2013)}]{smallenburg2013liquids}%
  \BibitemOpen
  \bibfield  {author} {\bibinfo {author} {\bibfnamefont {F.}~\bibnamefont
  {Smallenburg}}\ and\ \bibinfo {author} {\bibfnamefont {F.}~\bibnamefont
  {Sciortino}},\ }\href@noop {} {\bibfield  {journal} {\bibinfo  {journal}
  {Nat. Phys.}\ }\textbf {\bibinfo {volume} {9}},\ \bibinfo {pages} {554}
  (\bibinfo {year} {2013})}\BibitemShut {NoStop}%
\bibitem [{\citenamefont {Rold{\'a}n-Vargas}\ \emph {et~al.}(2013)\citenamefont
  {Rold{\'a}n-Vargas}, \citenamefont {Smallenburg}, \citenamefont {Kob},\ and\
  \citenamefont {Sciortino}}]{roldan2013gelling}%
  \BibitemOpen
  \bibfield  {author} {\bibinfo {author} {\bibfnamefont {S.}~\bibnamefont
  {Rold{\'a}n-Vargas}}, \bibinfo {author} {\bibfnamefont {F.}~\bibnamefont
  {Smallenburg}}, \bibinfo {author} {\bibfnamefont {W.}~\bibnamefont {Kob}}, \
  and\ \bibinfo {author} {\bibfnamefont {F.}~\bibnamefont {Sciortino}},\
  }\href@noop {} {\bibfield  {journal} {\bibinfo  {journal} {Sci. Rep.}\
  }\textbf {\bibinfo {volume} {3}} (\bibinfo {year} {2013})}\BibitemShut
  {NoStop}%
\bibitem [{\citenamefont {Duguet}\ \emph {et~al.}(2016)\citenamefont {Duguet},
  \citenamefont {Hubert}, \citenamefont {Chomette}, \citenamefont {Perro},\
  and\ \citenamefont {Ravaine}}]{duguet2016patchy}%
  \BibitemOpen
  \bibfield  {author} {\bibinfo {author} {\bibfnamefont {{\'E}.}~\bibnamefont
  {Duguet}}, \bibinfo {author} {\bibfnamefont {C.}~\bibnamefont {Hubert}},
  \bibinfo {author} {\bibfnamefont {C.}~\bibnamefont {Chomette}}, \bibinfo
  {author} {\bibfnamefont {A.}~\bibnamefont {Perro}}, \ and\ \bibinfo {author}
  {\bibfnamefont {S.}~\bibnamefont {Ravaine}},\ }\href@noop {} {\bibfield
  {journal} {\bibinfo  {journal} {Comptes Rendus Chimie}\ }\textbf {\bibinfo
  {volume} {19}},\ \bibinfo {pages} {173} (\bibinfo {year} {2016})}\BibitemShut
  {NoStop}%
\bibitem [{\citenamefont {Bianchi}\ \emph {et~al.}(2006)\citenamefont
  {Bianchi}, \citenamefont {Largo}, \citenamefont {Tartaglia}, \citenamefont
  {Zaccarelli},\ and\ \citenamefont {Sciortino}}]{bianchi2006phase}%
  \BibitemOpen
  \bibfield  {author} {\bibinfo {author} {\bibfnamefont {E.}~\bibnamefont
  {Bianchi}}, \bibinfo {author} {\bibfnamefont {J.}~\bibnamefont {Largo}},
  \bibinfo {author} {\bibfnamefont {P.}~\bibnamefont {Tartaglia}}, \bibinfo
  {author} {\bibfnamefont {E.}~\bibnamefont {Zaccarelli}}, \ and\ \bibinfo
  {author} {\bibfnamefont {F.}~\bibnamefont {Sciortino}},\ }\href@noop {}
  {\bibfield  {journal} {\bibinfo  {journal} {Phys. Rev. Lett.}\ }\textbf
  {\bibinfo {volume} {97}},\ \bibinfo {pages} {168301} (\bibinfo {year}
  {2006})}\BibitemShut {NoStop}%
\bibitem [{\citenamefont {Sciortino}\ and\ \citenamefont
  {Zaccarelli}(2017)}]{sciortino2017equilibrium}%
  \BibitemOpen
  \bibfield  {author} {\bibinfo {author} {\bibfnamefont {F.}~\bibnamefont
  {Sciortino}}\ and\ \bibinfo {author} {\bibfnamefont {E.}~\bibnamefont
  {Zaccarelli}},\ }\href@noop {} {\bibfield  {journal} {\bibinfo  {journal}
  {Curr. Opin. Colloid. In.}\ } (\bibinfo {year} {2017})}\BibitemShut {NoStop}%
\bibitem [{\citenamefont {Cai}\ \emph {et~al.}(2017)\citenamefont {Cai},
  \citenamefont {Townsend}, \citenamefont {Dodson}, \citenamefont {Heiney},\
  and\ \citenamefont {Sweeney}}]{cai2017eye}%
  \BibitemOpen
  \bibfield  {author} {\bibinfo {author} {\bibfnamefont {J.}~\bibnamefont
  {Cai}}, \bibinfo {author} {\bibfnamefont {J.}~\bibnamefont {Townsend}},
  \bibinfo {author} {\bibfnamefont {T.}~\bibnamefont {Dodson}}, \bibinfo
  {author} {\bibfnamefont {P.}~\bibnamefont {Heiney}}, \ and\ \bibinfo {author}
  {\bibfnamefont {A.}~\bibnamefont {Sweeney}},\ }\href@noop {} {\bibfield
  {journal} {\bibinfo  {journal} {Science}\ }\textbf {\bibinfo {volume}
  {357}},\ \bibinfo {pages} {564} (\bibinfo {year} {2017})}\BibitemShut
  {NoStop}%
\bibitem [{\citenamefont {Bianchi}\ \emph {et~al.}(2017)\citenamefont
  {Bianchi}, \citenamefont {Capone}, \citenamefont {Coluzza}, \citenamefont
  {Rovigatti},\ and\ \citenamefont {van Oostrum}}]{C7CP03149A}%
  \BibitemOpen
  \bibfield  {author} {\bibinfo {author} {\bibfnamefont {E.}~\bibnamefont
  {Bianchi}}, \bibinfo {author} {\bibfnamefont {B.}~\bibnamefont {Capone}},
  \bibinfo {author} {\bibfnamefont {I.}~\bibnamefont {Coluzza}}, \bibinfo
  {author} {\bibfnamefont {L.}~\bibnamefont {Rovigatti}}, \ and\ \bibinfo
  {author} {\bibfnamefont {P.~D.~J.}\ \bibnamefont {van Oostrum}},\ }\href
  {\doibase 10.1039/C7CP03149A} {\bibfield  {journal} {\bibinfo  {journal}
  {Phys. Chem. Chem. Phys.}\ }\textbf {\bibinfo {volume} {19}},\ \bibinfo
  {pages} {19847} (\bibinfo {year} {2017})}\BibitemShut {NoStop}%
\bibitem [{\citenamefont {Sakai}\ \emph {et~al.}(2008)\citenamefont {Sakai},
  \citenamefont {Matsunaga}, \citenamefont {Yamamoto}, \citenamefont {Ito},
  \citenamefont {Yoshida}, \citenamefont {Suzuki}, \citenamefont {Sasaki},
  \citenamefont {Shibayama},\ and\ \citenamefont {Chung}}]{sakai2008design}%
  \BibitemOpen
  \bibfield  {author} {\bibinfo {author} {\bibfnamefont {T.}~\bibnamefont
  {Sakai}}, \bibinfo {author} {\bibfnamefont {T.}~\bibnamefont {Matsunaga}},
  \bibinfo {author} {\bibfnamefont {Y.}~\bibnamefont {Yamamoto}}, \bibinfo
  {author} {\bibfnamefont {C.}~\bibnamefont {Ito}}, \bibinfo {author}
  {\bibfnamefont {R.}~\bibnamefont {Yoshida}}, \bibinfo {author} {\bibfnamefont
  {S.}~\bibnamefont {Suzuki}}, \bibinfo {author} {\bibfnamefont
  {N.}~\bibnamefont {Sasaki}}, \bibinfo {author} {\bibfnamefont
  {M.}~\bibnamefont {Shibayama}}, \ and\ \bibinfo {author} {\bibfnamefont
  {U.-i.}\ \bibnamefont {Chung}},\ }\href@noop {} {\bibfield  {journal}
  {\bibinfo  {journal} {Macromolecules}\ }\textbf {\bibinfo {volume} {41}},\
  \bibinfo {pages} {5379} (\bibinfo {year} {2008})}\BibitemShut {NoStop}%
\bibitem [{\citenamefont {Li}\ \emph {et~al.}(2017)\citenamefont {Li},
  \citenamefont {Hirosawa}, \citenamefont {Sakai}, \citenamefont {Gilbert},\
  and\ \citenamefont {Shibayama}}]{li2017sans}%
  \BibitemOpen
  \bibfield  {author} {\bibinfo {author} {\bibfnamefont {X.}~\bibnamefont
  {Li}}, \bibinfo {author} {\bibfnamefont {K.}~\bibnamefont {Hirosawa}},
  \bibinfo {author} {\bibfnamefont {T.}~\bibnamefont {Sakai}}, \bibinfo
  {author} {\bibfnamefont {E.~P.}\ \bibnamefont {Gilbert}}, \ and\ \bibinfo
  {author} {\bibfnamefont {M.}~\bibnamefont {Shibayama}},\ }\href@noop {}
  {\bibfield  {journal} {\bibinfo  {journal} {Macromolecules}\ }\textbf
  {\bibinfo {volume} {50}},\ \bibinfo {pages} {3655} (\bibinfo {year}
  {2017})}\BibitemShut {NoStop}%
\bibitem [{\citenamefont {Bomboi}\ \emph {et~al.}(2015)\citenamefont {Bomboi},
  \citenamefont {Biffi}, \citenamefont {Cerbino}, \citenamefont {Bellini},
  \citenamefont {Bordi},\ and\ \citenamefont
  {Sciortino}}]{bomboi2015equilibrium}%
  \BibitemOpen
  \bibfield  {author} {\bibinfo {author} {\bibfnamefont {F.}~\bibnamefont
  {Bomboi}}, \bibinfo {author} {\bibfnamefont {S.}~\bibnamefont {Biffi}},
  \bibinfo {author} {\bibfnamefont {R.}~\bibnamefont {Cerbino}}, \bibinfo
  {author} {\bibfnamefont {T.}~\bibnamefont {Bellini}}, \bibinfo {author}
  {\bibfnamefont {F.}~\bibnamefont {Bordi}}, \ and\ \bibinfo {author}
  {\bibfnamefont {F.}~\bibnamefont {Sciortino}},\ }\href@noop {} {\bibfield
  {journal} {\bibinfo  {journal} {Eur. Phys. J. E}\ }\textbf {\bibinfo {volume}
  {38}},\ \bibinfo {pages} {1} (\bibinfo {year} {2015})}\BibitemShut {NoStop}%
\bibitem [{\citenamefont {Romano}\ and\ \citenamefont
  {Sciortino}(2015)}]{romano2015switching}%
  \BibitemOpen
  \bibfield  {author} {\bibinfo {author} {\bibfnamefont {F.}~\bibnamefont
  {Romano}}\ and\ \bibinfo {author} {\bibfnamefont {F.}~\bibnamefont
  {Sciortino}},\ }\href@noop {} {\bibfield  {journal} {\bibinfo  {journal}
  {Phys. Rev. Lett.}\ }\textbf {\bibinfo {volume} {114}},\ \bibinfo {pages}
  {078104} (\bibinfo {year} {2015})}\BibitemShut {NoStop}%
\bibitem [{\citenamefont {Montarnal}\ \emph {et~al.}(2011)\citenamefont
  {Montarnal}, \citenamefont {Capelot}, \citenamefont {Tournilhac},\ and\
  \citenamefont {Leibler}}]{montarnal2011silica}%
  \BibitemOpen
  \bibfield  {author} {\bibinfo {author} {\bibfnamefont {D.}~\bibnamefont
  {Montarnal}}, \bibinfo {author} {\bibfnamefont {M.}~\bibnamefont {Capelot}},
  \bibinfo {author} {\bibfnamefont {F.}~\bibnamefont {Tournilhac}}, \ and\
  \bibinfo {author} {\bibfnamefont {L.}~\bibnamefont {Leibler}},\ }\href@noop
  {} {\bibfield  {journal} {\bibinfo  {journal} {Science}\ }\textbf {\bibinfo
  {volume} {334}},\ \bibinfo {pages} {965} (\bibinfo {year}
  {2011})}\BibitemShut {NoStop}%
\bibitem [{\citenamefont {Denissen}, \citenamefont {Winne},\ and\ \citenamefont
  {Du~Prez}(2016)}]{denissen2016vitrimers}%
  \BibitemOpen
  \bibfield  {author} {\bibinfo {author} {\bibfnamefont {W.}~\bibnamefont
  {Denissen}}, \bibinfo {author} {\bibfnamefont {J.~M.}\ \bibnamefont {Winne}},
  \ and\ \bibinfo {author} {\bibfnamefont {F.~E.}\ \bibnamefont {Du~Prez}},\
  }\href@noop {} {\bibfield  {journal} {\bibinfo  {journal} {Chem. Sci.}\
  }\textbf {\bibinfo {volume} {7}},\ \bibinfo {pages} {30} (\bibinfo {year}
  {2016})}\BibitemShut {NoStop}%
\bibitem [{\citenamefont {Gnan}, \citenamefont {Zaccarelli},\ and\
  \citenamefont {Sciortino}(2014)}]{gnancasimir2014}%
  \BibitemOpen
  \bibfield  {author} {\bibinfo {author} {\bibfnamefont {N.}~\bibnamefont
  {Gnan}}, \bibinfo {author} {\bibfnamefont {E.}~\bibnamefont {Zaccarelli}}, \
  and\ \bibinfo {author} {\bibfnamefont {F.}~\bibnamefont {Sciortino}},\
  }\href@noop {} {\bibfield  {journal} {\bibinfo  {journal} {Nat. Comm.}\
  }\textbf {\bibinfo {volume} {5}},\ \bibinfo {pages} {3267} (\bibinfo {year}
  {2014})}\BibitemShut {NoStop}%
\bibitem [{\citenamefont {Gnan}\ \emph {et~al.}(2012)\citenamefont {Gnan},
  \citenamefont {Zaccarelli}, \citenamefont {Tartaglia},\ and\ \citenamefont
  {Sciortino}}]{gnan2012properties}%
  \BibitemOpen
  \bibfield  {author} {\bibinfo {author} {\bibfnamefont {N.}~\bibnamefont
  {Gnan}}, \bibinfo {author} {\bibfnamefont {E.}~\bibnamefont {Zaccarelli}},
  \bibinfo {author} {\bibfnamefont {P.}~\bibnamefont {Tartaglia}}, \ and\
  \bibinfo {author} {\bibfnamefont {F.}~\bibnamefont {Sciortino}},\ }\href@noop
  {} {\bibfield  {journal} {\bibinfo  {journal} {Soft Matter}\ }\textbf
  {\bibinfo {volume} {8}},\ \bibinfo {pages} {1991} (\bibinfo {year}
  {2012})}\BibitemShut {NoStop}%
\bibitem [{\citenamefont {Flory}(1941)}]{flory1941molecular}%
  \BibitemOpen
  \bibfield  {author} {\bibinfo {author} {\bibfnamefont {P.~J.}\ \bibnamefont
  {Flory}},\ }\href@noop {} {\bibfield  {journal} {\bibinfo  {journal} {J. Am.
  Chem. Soc.}\ }\textbf {\bibinfo {volume} {63}},\ \bibinfo {pages} {3083}
  (\bibinfo {year} {1941})}\BibitemShut {NoStop}%
\bibitem [{\citenamefont {Stockmayer}(1943)}]{stockmayer1943theory}%
  \BibitemOpen
  \bibfield  {author} {\bibinfo {author} {\bibfnamefont {W.~H.}\ \bibnamefont
  {Stockmayer}},\ }\href@noop {} {\bibfield  {journal} {\bibinfo  {journal} {J.
  Chem. Phys.}\ }\textbf {\bibinfo {volume} {11}},\ \bibinfo {pages} {45}
  (\bibinfo {year} {1943})}\BibitemShut {NoStop}%
\bibitem [{\citenamefont {Liu}\ and\ \citenamefont
  {West}(2004)}]{liu2004happy}%
  \BibitemOpen
  \bibfield  {author} {\bibinfo {author} {\bibfnamefont {Y.}~\bibnamefont
  {Liu}}\ and\ \bibinfo {author} {\bibfnamefont {S.~C.}\ \bibnamefont {West}},\
  }\href@noop {} {\bibfield  {journal} {\bibinfo  {journal} {Nat. Rev. Mol.
  Cell Biol.}\ }\textbf {\bibinfo {volume} {5}},\ \bibinfo {pages} {937}
  (\bibinfo {year} {2004})}\BibitemShut {NoStop}%
\bibitem [{\citenamefont {Wang}\ \emph {et~al.}(2016)\citenamefont {Wang},
  \citenamefont {Nocka}, \citenamefont {Wiemann}, \citenamefont {Hinckley},
  \citenamefont {Mukerji},\ and\ \citenamefont {Starr}}]{wang2016holliday}%
  \BibitemOpen
  \bibfield  {author} {\bibinfo {author} {\bibfnamefont {W.}~\bibnamefont
  {Wang}}, \bibinfo {author} {\bibfnamefont {L.~M.}\ \bibnamefont {Nocka}},
  \bibinfo {author} {\bibfnamefont {B.~Z.}\ \bibnamefont {Wiemann}}, \bibinfo
  {author} {\bibfnamefont {D.~M.}\ \bibnamefont {Hinckley}}, \bibinfo {author}
  {\bibfnamefont {I.}~\bibnamefont {Mukerji}}, \ and\ \bibinfo {author}
  {\bibfnamefont {F.~W.}\ \bibnamefont {Starr}},\ }\href@noop {} {\bibfield
  {journal} {\bibinfo  {journal} {Sci. Rep.}\ }\textbf {\bibinfo {volume}
  {6}},\ \bibinfo {pages} {22863} (\bibinfo {year} {2016})}\BibitemShut
  {NoStop}%
\bibitem [{\citenamefont {van Gool}\ \emph {et~al.}(1999)\citenamefont {van
  Gool}, \citenamefont {Hajibagheri}, \citenamefont {Stasiak},\ and\
  \citenamefont {West}}]{van1999assembly}%
  \BibitemOpen
  \bibfield  {author} {\bibinfo {author} {\bibfnamefont {A.~J.}\ \bibnamefont
  {van Gool}}, \bibinfo {author} {\bibfnamefont {N.~M.}\ \bibnamefont
  {Hajibagheri}}, \bibinfo {author} {\bibfnamefont {A.}~\bibnamefont
  {Stasiak}}, \ and\ \bibinfo {author} {\bibfnamefont {S.~C.}\ \bibnamefont
  {West}},\ }\href@noop {} {\bibfield  {journal} {\bibinfo  {journal} {Genes
  Dev.}\ }\textbf {\bibinfo {volume} {13}},\ \bibinfo {pages} {1861} (\bibinfo
  {year} {1999})}\BibitemShut {NoStop}%
\bibitem [{\citenamefont {Sharples}, \citenamefont {Ingleston},\ and\
  \citenamefont {Lloyd}(1999)}]{sharples1999holliday}%
  \BibitemOpen
  \bibfield  {author} {\bibinfo {author} {\bibfnamefont {G.~J.}\ \bibnamefont
  {Sharples}}, \bibinfo {author} {\bibfnamefont {S.~M.}\ \bibnamefont
  {Ingleston}}, \ and\ \bibinfo {author} {\bibfnamefont {R.~G.}\ \bibnamefont
  {Lloyd}},\ }\href@noop {} {\bibfield  {journal} {\bibinfo  {journal} {J.
  Bacteriol.}\ }\textbf {\bibinfo {volume} {181}},\ \bibinfo {pages} {5543}
  (\bibinfo {year} {1999})}\BibitemShut {NoStop}%
\bibitem [{\citenamefont {Seeman}(2003{\natexlab{a}})}]{seeman2003crossroads}%
  \BibitemOpen
  \bibfield  {author} {\bibinfo {author} {\bibfnamefont {N.~C.}\ \bibnamefont
  {Seeman}},\ }\href@noop {} {\bibfield  {journal} {\bibinfo  {journal} {Chem.
  Biol.}\ }\textbf {\bibinfo {volume} {10}},\ \bibinfo {pages} {1151} (\bibinfo
  {year} {2003}{\natexlab{a}})}\BibitemShut {NoStop}%
\bibitem [{\citenamefont {Seeman}(2003{\natexlab{b}})}]{seeman2003dna}%
  \BibitemOpen
  \bibfield  {author} {\bibinfo {author} {\bibfnamefont {N.~C.}\ \bibnamefont
  {Seeman}},\ }\href@noop {} {\bibfield  {journal} {\bibinfo  {journal}
  {Nature}\ }\textbf {\bibinfo {volume} {421}},\ \bibinfo {pages} {427}
  (\bibinfo {year} {2003}{\natexlab{b}})}\BibitemShut {NoStop}%
\bibitem [{\citenamefont {Li}\ \emph {et~al.}(2004)\citenamefont {Li},
  \citenamefont {Tseng}, \citenamefont {Kwon}, \citenamefont {d'Espaux},
  \citenamefont {Bunch}, \citenamefont {McEuen},\ and\ \citenamefont
  {Luo}}]{li2004controlled}%
  \BibitemOpen
  \bibfield  {author} {\bibinfo {author} {\bibfnamefont {Y.}~\bibnamefont
  {Li}}, \bibinfo {author} {\bibfnamefont {Y.~D.}\ \bibnamefont {Tseng}},
  \bibinfo {author} {\bibfnamefont {S.~Y.}\ \bibnamefont {Kwon}}, \bibinfo
  {author} {\bibfnamefont {L.}~\bibnamefont {d'Espaux}}, \bibinfo {author}
  {\bibfnamefont {J.~S.}\ \bibnamefont {Bunch}}, \bibinfo {author}
  {\bibfnamefont {P.~L.}\ \bibnamefont {McEuen}}, \ and\ \bibinfo {author}
  {\bibfnamefont {D.}~\bibnamefont {Luo}},\ }\href@noop {} {\bibfield
  {journal} {\bibinfo  {journal} {Nat. Mater.}\ }\textbf {\bibinfo {volume}
  {3}},\ \bibinfo {pages} {38} (\bibinfo {year} {2004})}\BibitemShut {NoStop}%
\bibitem [{\citenamefont {Flory}(1953)}]{flory1953principles}%
  \BibitemOpen
  \bibfield  {author} {\bibinfo {author} {\bibfnamefont {P.~J.}\ \bibnamefont
  {Flory}},\ }\href@noop {} {\emph {\bibinfo {title} {Principles of polymer
  chemistry}}}\ (\bibinfo  {publisher} {Cornell University Press},\ \bibinfo
  {year} {1953})\BibitemShut {NoStop}%
\bibitem [{\citenamefont {{P. Desjardins, and D.
  Conklin}}(2010)}]{2010_Nanodrop}%
  \BibitemOpen
  \bibfield  {author} {\bibinfo {author} {\bibnamefont {{P. Desjardins, and D.
  Conklin}}},\ }\href@noop {} {\bibfield  {journal} {\bibinfo  {journal} {{J.
  Vis. Exp.}}\ }\textbf {\bibinfo {volume} {45}} (\bibinfo {year}
  {2010})}\BibitemShut {NoStop}%
\bibitem [{\citenamefont {Wertheim}(1984{\natexlab{a}})}]{wertheim1984fluids1}%
  \BibitemOpen
  \bibfield  {author} {\bibinfo {author} {\bibfnamefont {M.}~\bibnamefont
  {Wertheim}},\ }\href@noop {} {\bibfield  {journal} {\bibinfo  {journal} {J.
  Stat. Phys.}\ }\textbf {\bibinfo {volume} {35}},\ \bibinfo {pages} {19}
  (\bibinfo {year} {1984}{\natexlab{a}})}\BibitemShut {NoStop}%
\bibitem [{\citenamefont {Wertheim}(1984{\natexlab{b}})}]{wertheim1984fluids2}%
  \BibitemOpen
  \bibfield  {author} {\bibinfo {author} {\bibfnamefont {M.}~\bibnamefont
  {Wertheim}},\ }\href@noop {} {\bibfield  {journal} {\bibinfo  {journal} {J.
  Stat. Phys.}\ }\textbf {\bibinfo {volume} {35}},\ \bibinfo {pages} {35}
  (\bibinfo {year} {1984}{\natexlab{b}})}\BibitemShut {NoStop}%
\bibitem [{\citenamefont {Smallenburg}, \citenamefont {Leibler},\ and\
  \citenamefont {Sciortino}(2013)}]{smallenburg2013patchy}%
  \BibitemOpen
  \bibfield  {author} {\bibinfo {author} {\bibfnamefont {F.}~\bibnamefont
  {Smallenburg}}, \bibinfo {author} {\bibfnamefont {L.}~\bibnamefont
  {Leibler}}, \ and\ \bibinfo {author} {\bibfnamefont {F.}~\bibnamefont
  {Sciortino}},\ }\href@noop {} {\bibfield  {journal} {\bibinfo  {journal}
  {Phys. Rev. Lett.}\ }\textbf {\bibinfo {volume} {111}},\ \bibinfo {pages}
  {188002} (\bibinfo {year} {2013})}\BibitemShut {NoStop}%
\bibitem [{\citenamefont {Berne}\ and\ \citenamefont
  {Pecora}(2000)}]{berne2000dynamic}%
  \BibitemOpen
  \bibfield  {author} {\bibinfo {author} {\bibfnamefont {B.~J.}\ \bibnamefont
  {Berne}}\ and\ \bibinfo {author} {\bibfnamefont {R.}~\bibnamefont {Pecora}},\
  }\href@noop {} {\emph {\bibinfo {title} {Dynamic light scattering: with
  applications to chemistry, biology, and physics}}}\ (\bibinfo  {publisher}
  {Courier Corporation},\ \bibinfo {year} {2000})\BibitemShut {NoStop}%
\bibitem [{\citenamefont {Stockmayer}(1952)}]{stockmayer1952molecular}%
  \BibitemOpen
  \bibfield  {author} {\bibinfo {author} {\bibfnamefont {W.~H.}\ \bibnamefont
  {Stockmayer}},\ }\href@noop {} {\bibfield  {journal} {\bibinfo  {journal} {J.
  Polym. Sci. A1}\ }\textbf {\bibinfo {volume} {9}},\ \bibinfo {pages} {69}
  (\bibinfo {year} {1952})}\BibitemShut {NoStop}%
\bibitem [{\citenamefont {SantaLucia}(1998)}]{SantaLucia17021998}%
  \BibitemOpen
  \bibfield  {author} {\bibinfo {author} {\bibfnamefont {J.}~\bibnamefont
  {SantaLucia}},\ }\href@noop {} {\bibfield  {journal} {\bibinfo  {journal}
  {Proc. Natl. Acad. Sci.}\ }\textbf {\bibinfo {volume} {95}},\ \bibinfo
  {pages} {1460} (\bibinfo {year} {1998})}\BibitemShut {NoStop}%
\bibitem [{nup()}]{nupack}%
  \BibitemOpen
  \href@noop {} {\bibinfo  {journal} {http://www.nupack.org/}\ }\BibitemShut
  {NoStop}%
\bibitem [{\citenamefont {Huang}(2009)}]{huang2009introduction}%
  \BibitemOpen
\bibfield  {journal} {  }\bibfield  {author} {\bibinfo {author} {\bibfnamefont
  {K.}~\bibnamefont {Huang}},\ }\href@noop {} {\emph {\bibinfo {title}
  {Introduction to statistical physics}}}\ (\bibinfo  {publisher} {CRC press},\
  \bibinfo {year} {2009})\BibitemShut {NoStop}%
\bibitem [{\citenamefont {Rubinstein}\ and\ \citenamefont
  {Colby}(2003)}]{rubinstein2003polymer}%
  \BibitemOpen
  \bibfield  {author} {\bibinfo {author} {\bibfnamefont {M.}~\bibnamefont
  {Rubinstein}}\ and\ \bibinfo {author} {\bibfnamefont {R.~H.}\ \bibnamefont
  {Colby}},\ }\href@noop {} {\emph {\bibinfo {title} {Polymer physics}}},\
  Vol.~\bibinfo {volume} {23}\ (\bibinfo  {publisher} {Oxford University Press
  New York},\ \bibinfo {year} {2003})\BibitemShut {NoStop}%
\bibitem [{\citenamefont {Russo}, \citenamefont {Tartaglia},\ and\
  \citenamefont {Sciortino}(2009)}]{russo2009reversible}%
  \BibitemOpen
  \bibfield  {author} {\bibinfo {author} {\bibfnamefont {J.}~\bibnamefont
  {Russo}}, \bibinfo {author} {\bibfnamefont {P.}~\bibnamefont {Tartaglia}}, \
  and\ \bibinfo {author} {\bibfnamefont {F.}~\bibnamefont {Sciortino}},\
  }\href@noop {} {\bibfield  {journal} {\bibinfo  {journal} {J. Chem. Phys.}\
  }\textbf {\bibinfo {volume} {131}},\ \bibinfo {pages} {014504} (\bibinfo
  {year} {2009})}\BibitemShut {NoStop}%
\bibitem [{\citenamefont {{\v{S}}poner}, \citenamefont {Leszczynski},\ and\
  \citenamefont {Hobza}(1996)}]{vsponer1996structures}%
  \BibitemOpen
  \bibfield  {author} {\bibinfo {author} {\bibfnamefont {J.}~\bibnamefont
  {{\v{S}}poner}}, \bibinfo {author} {\bibfnamefont {J.}~\bibnamefont
  {Leszczynski}}, \ and\ \bibinfo {author} {\bibfnamefont {P.}~\bibnamefont
  {Hobza}},\ }\href@noop {} {\bibfield  {journal} {\bibinfo  {journal} {J.
  Chem. Phys.}\ }\textbf {\bibinfo {volume} {100}},\ \bibinfo {pages} {1965}
  (\bibinfo {year} {1996})}\BibitemShut {NoStop}%
\bibitem [{\citenamefont {Schumakovitch}\ \emph {et~al.}(2002)\citenamefont
  {Schumakovitch}, \citenamefont {Grange}, \citenamefont {Strunz},
  \citenamefont {Bertoncini}, \citenamefont {G{\"u}ntherodt},\ and\
  \citenamefont {Hegner}}]{schumakovitch2002temperature}%
  \BibitemOpen
  \bibfield  {author} {\bibinfo {author} {\bibfnamefont {I.}~\bibnamefont
  {Schumakovitch}}, \bibinfo {author} {\bibfnamefont {W.}~\bibnamefont
  {Grange}}, \bibinfo {author} {\bibfnamefont {T.}~\bibnamefont {Strunz}},
  \bibinfo {author} {\bibfnamefont {P.}~\bibnamefont {Bertoncini}}, \bibinfo
  {author} {\bibfnamefont {H.-J.}\ \bibnamefont {G{\"u}ntherodt}}, \ and\
  \bibinfo {author} {\bibfnamefont {M.}~\bibnamefont {Hegner}},\ }\href@noop {}
  {\bibfield  {journal} {\bibinfo  {journal} {Biophys. J.}\ }\textbf {\bibinfo
  {volume} {82}},\ \bibinfo {pages} {517} (\bibinfo {year} {2002})}\BibitemShut
  {NoStop}%
\bibitem [{\citenamefont {Henning{\'a}~Winter}\ \emph
  {et~al.}(1995)\citenamefont {Henning{\'a}~Winter} \emph
  {et~al.}}]{henningawinter1995rheological}%
  \BibitemOpen
  \bibfield  {author} {\bibinfo {author} {\bibfnamefont {H.}~\bibnamefont
  {Henning{\'a}~Winter}} \emph {et~al.},\ }\href@noop {} {\bibfield  {journal}
  {\bibinfo  {journal} {Faraday Discuss.}\ }\textbf {\bibinfo {volume} {101}},\
  \bibinfo {pages} {93} (\bibinfo {year} {1995})}\BibitemShut {NoStop}%
\bibitem [{\citenamefont {Lindquist}\ \emph {et~al.}(2016)\citenamefont
  {Lindquist}, \citenamefont {Jadrich}, \citenamefont {Milliron},\ and\
  \citenamefont {Truskett}}]{lindquist2016formation}%
  \BibitemOpen
  \bibfield  {author} {\bibinfo {author} {\bibfnamefont {B.~A.}\ \bibnamefont
  {Lindquist}}, \bibinfo {author} {\bibfnamefont {R.~B.}\ \bibnamefont
  {Jadrich}}, \bibinfo {author} {\bibfnamefont {D.~J.}\ \bibnamefont
  {Milliron}}, \ and\ \bibinfo {author} {\bibfnamefont {T.~M.}\ \bibnamefont
  {Truskett}},\ }\href@noop {} {\bibfield  {journal} {\bibinfo  {journal} {J.
  Chem. Phys.}\ }\textbf {\bibinfo {volume} {145}},\ \bibinfo {pages} {074906}
  (\bibinfo {year} {2016})}\BibitemShut {NoStop}%
\bibitem [{\citenamefont {Adam}\ \emph {et~al.}(1988)\citenamefont {Adam},
  \citenamefont {Delsanti}, \citenamefont {Munch},\ and\ \citenamefont
  {Durand}}]{adam1988dynamical}%
  \BibitemOpen
  \bibfield  {author} {\bibinfo {author} {\bibfnamefont {M.}~\bibnamefont
  {Adam}}, \bibinfo {author} {\bibfnamefont {M.}~\bibnamefont {Delsanti}},
  \bibinfo {author} {\bibfnamefont {J.}~\bibnamefont {Munch}}, \ and\ \bibinfo
  {author} {\bibfnamefont {D.}~\bibnamefont {Durand}},\ }\href@noop {}
  {\bibfield  {journal} {\bibinfo  {journal} {Phys. Rev. Lett.}\ }\textbf
  {\bibinfo {volume} {61}},\ \bibinfo {pages} {706} (\bibinfo {year}
  {1988})}\BibitemShut {NoStop}%
\bibitem [{\citenamefont {Richter}(2007)}]{richter2007comparison}%
  \BibitemOpen
  \bibfield  {author} {\bibinfo {author} {\bibfnamefont {S.}~\bibnamefont
  {Richter}},\ }\href@noop {} {\bibfield  {journal} {\bibinfo  {journal}
  {Macromol. Symp.}\ }\textbf {\bibinfo {volume} {256}},\ \bibinfo {pages} {88}
  (\bibinfo {year} {2007})}\BibitemShut {NoStop}%
\bibitem [{\citenamefont {Berthon}\ \emph {et~al.}(2001)\citenamefont
  {Berthon}, \citenamefont {Barbieri}, \citenamefont {Ehrburger-Dolle},
  \citenamefont {Geissler}, \citenamefont {Achard}, \citenamefont {Bley},
  \citenamefont {Hecht}, \citenamefont {Livet}, \citenamefont {Pajonk},
  \citenamefont {Pinto} \emph {et~al.}}]{berthon2001dls}%
  \BibitemOpen
  \bibfield  {author} {\bibinfo {author} {\bibfnamefont {S.}~\bibnamefont
  {Berthon}}, \bibinfo {author} {\bibfnamefont {O.}~\bibnamefont {Barbieri}},
  \bibinfo {author} {\bibfnamefont {F.}~\bibnamefont {Ehrburger-Dolle}},
  \bibinfo {author} {\bibfnamefont {E.}~\bibnamefont {Geissler}}, \bibinfo
  {author} {\bibfnamefont {P.}~\bibnamefont {Achard}}, \bibinfo {author}
  {\bibfnamefont {F.}~\bibnamefont {Bley}}, \bibinfo {author} {\bibfnamefont
  {A.-M.}\ \bibnamefont {Hecht}}, \bibinfo {author} {\bibfnamefont
  {F.}~\bibnamefont {Livet}}, \bibinfo {author} {\bibfnamefont {G.~M.}\
  \bibnamefont {Pajonk}}, \bibinfo {author} {\bibfnamefont {N.}~\bibnamefont
  {Pinto}},  \emph {et~al.},\ }\href@noop {} {\bibfield  {journal} {\bibinfo
  {journal} {J. Non-Cryst. Solids}\ }\textbf {\bibinfo {volume} {285}},\
  \bibinfo {pages} {154} (\bibinfo {year} {2001})}\BibitemShut {NoStop}%
\bibitem [{\citenamefont {Djabourov}, \citenamefont {Lechaire},\ and\
  \citenamefont {Gaill}(1993)}]{djabourov1993structure}%
  \BibitemOpen
  \bibfield  {author} {\bibinfo {author} {\bibfnamefont {M.}~\bibnamefont
  {Djabourov}}, \bibinfo {author} {\bibfnamefont {J.-P.}\ \bibnamefont
  {Lechaire}}, \ and\ \bibinfo {author} {\bibfnamefont {F.}~\bibnamefont
  {Gaill}},\ }\href@noop {} {\bibfield  {journal} {\bibinfo  {journal}
  {Biorheology}\ }\textbf {\bibinfo {volume} {30}},\ \bibinfo {pages} {191}
  (\bibinfo {year} {1993})}\BibitemShut {NoStop}%
\bibitem [{\citenamefont {Bordi}\ \emph {et~al.}(2002)\citenamefont {Bordi},
  \citenamefont {Paradossi}, \citenamefont {Rinaldi},\ and\ \citenamefont
  {Ruzicka}}]{bordi2002chemical}%
  \BibitemOpen
  \bibfield  {author} {\bibinfo {author} {\bibfnamefont {F.}~\bibnamefont
  {Bordi}}, \bibinfo {author} {\bibfnamefont {G.}~\bibnamefont {Paradossi}},
  \bibinfo {author} {\bibfnamefont {C.}~\bibnamefont {Rinaldi}}, \ and\
  \bibinfo {author} {\bibfnamefont {B.}~\bibnamefont {Ruzicka}},\ }\href@noop
  {} {\bibfield  {journal} {\bibinfo  {journal} {Phys. A}\ }\textbf {\bibinfo
  {volume} {304}},\ \bibinfo {pages} {119} (\bibinfo {year}
  {2002})}\BibitemShut {NoStop}%
\bibitem [{\citenamefont {Sciortino}\ and\ \citenamefont
  {Tartaglia}(1997)}]{sciortino1997relaxation}%
  \BibitemOpen
  \bibfield  {author} {\bibinfo {author} {\bibfnamefont {F.}~\bibnamefont
  {Sciortino}}\ and\ \bibinfo {author} {\bibfnamefont {P.}~\bibnamefont
  {Tartaglia}},\ }\href@noop {} {\bibfield  {journal} {\bibinfo  {journal}
  {Phys. A}\ }\textbf {\bibinfo {volume} {236}},\ \bibinfo {pages} {140}
  (\bibinfo {year} {1997})}\BibitemShut {NoStop}%
\bibitem [{\citenamefont {Nava}\ \emph {et~al.}(2017)\citenamefont {Nava},
  \citenamefont {Rossi}, \citenamefont {Biffi}, \citenamefont {Sciortino},\
  and\ \citenamefont {Bellini}}]{nava2017fluctuating}%
  \BibitemOpen
  \bibfield  {author} {\bibinfo {author} {\bibfnamefont {G.}~\bibnamefont
  {Nava}}, \bibinfo {author} {\bibfnamefont {M.}~\bibnamefont {Rossi}},
  \bibinfo {author} {\bibfnamefont {S.}~\bibnamefont {Biffi}}, \bibinfo
  {author} {\bibfnamefont {F.}~\bibnamefont {Sciortino}}, \ and\ \bibinfo
  {author} {\bibfnamefont {T.}~\bibnamefont {Bellini}},\ }\href@noop {}
  {\bibfield  {journal} {\bibinfo  {journal} {Phys. Rev. Lett.}\ }\textbf
  {\bibinfo {volume} {119}},\ \bibinfo {pages} {078002} (\bibinfo {year}
  {2017})}\BibitemShut {NoStop}%
\bibitem [{\citenamefont {Fabbian}\ \emph {et~al.}(1999)\citenamefont
  {Fabbian}, \citenamefont {G{\"o}tze}, \citenamefont {Sciortino},
  \citenamefont {Tartaglia},\ and\ \citenamefont {Thiery}}]{fabbian1999ideal}%
  \BibitemOpen
  \bibfield  {author} {\bibinfo {author} {\bibfnamefont {L.}~\bibnamefont
  {Fabbian}}, \bibinfo {author} {\bibfnamefont {W.}~\bibnamefont {G{\"o}tze}},
  \bibinfo {author} {\bibfnamefont {F.}~\bibnamefont {Sciortino}}, \bibinfo
  {author} {\bibfnamefont {P.}~\bibnamefont {Tartaglia}}, \ and\ \bibinfo
  {author} {\bibfnamefont {F.}~\bibnamefont {Thiery}},\ }\href@noop {}
  {\bibfield  {journal} {\bibinfo  {journal} {Phys. Rev. E}\ }\textbf {\bibinfo
  {volume} {59}},\ \bibinfo {pages} {R1347} (\bibinfo {year}
  {1999})}\BibitemShut {NoStop}%
\bibitem [{\citenamefont {Sciortino}, \citenamefont {Tartaglia},\ and\
  \citenamefont {Zaccarelli}(2003)}]{sciortino2003evidence}%
  \BibitemOpen
  \bibfield  {author} {\bibinfo {author} {\bibfnamefont {F.}~\bibnamefont
  {Sciortino}}, \bibinfo {author} {\bibfnamefont {P.}~\bibnamefont
  {Tartaglia}}, \ and\ \bibinfo {author} {\bibfnamefont {E.}~\bibnamefont
  {Zaccarelli}},\ }\href@noop {} {\bibfield  {journal} {\bibinfo  {journal}
  {Phys. Rev. Lett.}\ }\textbf {\bibinfo {volume} {91}},\ \bibinfo {pages}
  {268301} (\bibinfo {year} {2003})}\BibitemShut {NoStop}%
\bibitem [{\citenamefont {Pham}\ \emph {et~al.}(2002)\citenamefont {Pham},
  \citenamefont {Puertas}, \citenamefont {Bergenholtz}, \citenamefont
  {Egelhaaf}, \citenamefont {Moussa{\i}d}, \citenamefont {Pusey}, \citenamefont
  {Schofield}, \citenamefont {Cates}, \citenamefont {Fuchs},\ and\
  \citenamefont {Poon}}]{pham2002multiple}%
  \BibitemOpen
  \bibfield  {author} {\bibinfo {author} {\bibfnamefont {K.~N.}\ \bibnamefont
  {Pham}}, \bibinfo {author} {\bibfnamefont {A.~M.}\ \bibnamefont {Puertas}},
  \bibinfo {author} {\bibfnamefont {J.}~\bibnamefont {Bergenholtz}}, \bibinfo
  {author} {\bibfnamefont {S.~U.}\ \bibnamefont {Egelhaaf}}, \bibinfo {author}
  {\bibfnamefont {A.}~\bibnamefont {Moussa{\i}d}}, \bibinfo {author}
  {\bibfnamefont {P.~N.}\ \bibnamefont {Pusey}}, \bibinfo {author}
  {\bibfnamefont {A.~B.}\ \bibnamefont {Schofield}}, \bibinfo {author}
  {\bibfnamefont {M.~E.}\ \bibnamefont {Cates}}, \bibinfo {author}
  {\bibfnamefont {M.}~\bibnamefont {Fuchs}}, \ and\ \bibinfo {author}
  {\bibfnamefont {W.~C.}\ \bibnamefont {Poon}},\ }\href@noop {} {\bibfield
  {journal} {\bibinfo  {journal} {Science}\ }\textbf {\bibinfo {volume}
  {296}},\ \bibinfo {pages} {104} (\bibinfo {year} {2002})}\BibitemShut
  {NoStop}%
\bibitem [{\citenamefont {G{\"o}tze}\ and\ \citenamefont
  {Sperl}(2004)}]{gotze2004nearly}%
  \BibitemOpen
  \bibfield  {author} {\bibinfo {author} {\bibfnamefont {W.}~\bibnamefont
  {G{\"o}tze}}\ and\ \bibinfo {author} {\bibfnamefont {M.}~\bibnamefont
  {Sperl}},\ }\href@noop {} {\bibfield  {journal} {\bibinfo  {journal} {Phys.
  Rev. Lett.}\ }\textbf {\bibinfo {volume} {92}},\ \bibinfo {pages} {105701}
  (\bibinfo {year} {2004})}\BibitemShut {NoStop}%
\bibitem [{\citenamefont {Moreno}\ and\ \citenamefont
  {Colmenero}(2006)}]{moreno2006anomalous}%
  \BibitemOpen
  \bibfield  {author} {\bibinfo {author} {\bibfnamefont {A.~J.}\ \bibnamefont
  {Moreno}}\ and\ \bibinfo {author} {\bibfnamefont {J.}~\bibnamefont
  {Colmenero}},\ }\href@noop {} {\bibfield  {journal} {\bibinfo  {journal}
  {Phys. Rev.E}\ }\textbf {\bibinfo {volume} {74}},\ \bibinfo {pages} {021409}
  (\bibinfo {year} {2006})}\BibitemShut {NoStop}%
\bibitem [{\citenamefont {Glaser}, \citenamefont {Hallatschek},\ and\
  \citenamefont {Kroy}(2008)}]{glaser2008dynamic}%
  \BibitemOpen
  \bibfield  {author} {\bibinfo {author} {\bibfnamefont {J.}~\bibnamefont
  {Glaser}}, \bibinfo {author} {\bibfnamefont {O.}~\bibnamefont {Hallatschek}},
  \ and\ \bibinfo {author} {\bibfnamefont {K.}~\bibnamefont {Kroy}},\
  }\href@noop {} {\bibfield  {journal} {\bibinfo  {journal} {Eur. Phys. J. E}\
  }\textbf {\bibinfo {volume} {26}},\ \bibinfo {pages} {123} (\bibinfo {year}
  {2008})}\BibitemShut {NoStop}%
\bibitem [{\citenamefont {Kern}\ and\ \citenamefont
  {Frenkel}(2003)}]{kern2003fluid}%
  \BibitemOpen
  \bibfield  {author} {\bibinfo {author} {\bibfnamefont {N.}~\bibnamefont
  {Kern}}\ and\ \bibinfo {author} {\bibfnamefont {D.}~\bibnamefont {Frenkel}},\
  }\href@noop {} {\bibfield  {journal} {\bibinfo  {journal} {J. Chem. Phys.}\
  }\textbf {\bibinfo {volume} {118}},\ \bibinfo {pages} {9882} (\bibinfo {year}
  {2003})}\BibitemShut {NoStop}%
\end{thebibliography}%

\end{document}